# A nonuniform popularity-similarity optimization (nPSO) model to efficiently generate realistic complex networks with communities

Alessandro Muscoloni[1] and Carlo Vittorio Cannistraci[1,2,*]

[1]Biomedical Cybernetics Group, Biotechnology Center (BIOTEC), Center for Molecular and Cellular Bioengineering (CMCB), Center for Systems Biology Dresden (CSBD), Department of Physics, Technische Universität Dresden, Tatzberg 47/49, 01307 Dresden, Germany
[2]Brain bio-inspired computation (BBC) lab, IRCCS Centro Neurolesi "Bonino Pulejo", Messina, Italy

*Corresponding author: Carlo Vittorio Cannistraci (kalokagathos.agon@gmail.com)

## Abstract

The hidden metric space behind complex network topologies is a fervid topic in current network science and the hyperbolic space is one of the most studied, because it seems associated to the structural organization of many real complex systems. The Popularity-Similarity-Optimization (PSO) model simulates how random geometric graphs grow in the hyperbolic space, reproducing strong clustering and scale-free degree distribution, however it misses to reproduce an important feature of real complex networks, which is the community organization.

The Geometrical-Preferential-Attachment (GPA) model was recently developed to confer to the PSO also a community structure, which is obtained by forcing different angular regions of the hyperbolic disk to have variable level of attractiveness. However, the number and size of the communities cannot be explicitly controlled in the GPA, which is a clear limitation for real applications.

Here, we introduce the nonuniform PSO (nPSO) model that – differently from GPA - forces heterogeneous angular node attractiveness by sampling the angular coordinates from a tailored nonuniform probability distribution, for instance a mixture of Gaussians. The nPSO differs from GPA in other three aspects: it allows to explicitly fix the number and size of communities; it allows to tune their mixing property through the network temperature; it is efficient to generate networks with high clustering. After several tests we propose the nPSO as a valid and efficient model to generate networks with communities in the hyperbolic space, which can be adopted as a realistic benchmark for different tasks such as community detection and link prediction.

# Introduction

In recent years the study of hidden geometrical spaces behind complex network topologies has led to several developments and, currently, the hyperbolic space seems to be one of the most appropriate in order to explain many of the structural features observed in real networks [1]–[8]. In 2012 Papadopoulos et al. [5] introduced the Popularity-Similarity-Optimization (PSO) model in order to describe how random geometric graphs grow in the hyperbolic space optimizing a trade-off between popularity and similarity. In this framework, the popularity of the nodes is represented by the radial coordinate in the hyperbolic disk, whereas the angular coordinates distance is the geometrical counterpart of the similarity between the nodes. Networks generated through the PSO model exhibit strong clustering and a scale-free degree distribution, two among the peculiar properties that usually characterize real-world topologies [9]–[11]. However, another important feature commonly observed is the community structure [12]–[14], which is lacking in the PSO model. The reason is that the nodes are arranged over the angular coordinate space according to a uniform distribution, therefore, since the connection probabilities are inversely proportional to the hyperbolic distances, there are not angular regions containing a cluster of spatially close nodes that are more densely connected between each other than with the rest of the network. This issue has been addressed in a following study by Zuev et al. [15], introducing the geometric preferential attachment (GPA). The GPA couples the latent hyperbolic network geometry with preferential attachment of nodes to this geometry in order to generate networks with strong clustering, scale-free degree distribution and a non-trivial community structure [15]. The main assumption of the GPA model and simultaneously the main innovation with respect to the PSO model is that the angular coordinate space is not equally attractive everywhere. Practically, the GPA is characterized by heterogeneous angular attractiveness: regions of different attractiveness are designed according to the rationale that the higher the attractiveness of a region the higher the probability that the nodes are placed in that angular section. Although this general idea can be implemented in several ways, a high-level description of the procedure presented in the study of Zuev et al. [15] is as follows (see Methods for details). For each new node entering in the network, a set of candidate positions is defined (angular coordinate sampled uniformly at random, radial coordinate mathematically fixed) and to every candidate position is assigned a probability depending on the number of nodes that would be "close" to the entering node if it were placed in that position. The probability is also function of a parameter of initial attractiveness, which can be used to tune the heterogeneity of the angular coordinate distribution. However, the GPA model does not allow – at least in the form in which it is currently proposed - to directly control in an *explicit* and *efficient* way the number and size of the communities, a property that instead might be interesting, for example, while proposing a community detection benchmark. Furthemore, the GPA model does not take into account the possibility to vary the network temperature. For this reason we here introduce a variation of the PSO model, which we call nonuniform PSO (nPSO) model, whose key aspects are the possibility of: i) fixing the number and size of communities; ii) tuning their mixing property through the network temperature; iii) efficiently producing also highly clustered realistic networks.

## Results and Discussion

The idea behind the nPSO is quite intuitive. The sampling of the angular coordinates from a uniform distribution - which is used by the standard PSO - can be generalized to sampling from any distribution with a desired shape. In particular, a nonuniform distribution would indicate the presence of regions with different levels of node attractiveness. In this study, for simplicity, we will concentrate on the Gaussian mixture distribution, which we consider (without loss of generality) suitable for describing how to build a nonuniform distributed sample of nodes along the angular coordinates of a hyperbolic disk, with communities that emerge in correspondence of the different Gaussians. However, we want to stress that our nPSO model is general and can be implemented considering any mixture of desired distributions from which to sample the angular coordinates of the nodes.

Although the parameters of the Gaussian mixture distribution built on the angular coordinate space allow for the investigation of disparate scenarios, in this investigation we focused on the most straightforward setting. For a given number of communities $C$, we consider a mixture of $C$ components with the means equidistantly arranged over the angular space, the same standard deviation and equal mixing proportions (see Methods for details). Fig. 1 reports examples of nonuniform distributions using 4 and 8 Gaussians that generate respectively the same number of communities. In this framework, the community membership is assigned considering for each node the mixing component whose mean is the closest in the angular space. Fig. 2 shows examples of networks in the hyperbolic space generated using the nPSO model for different values of clustering (temperature, $T = 0.1, 0.5, 0.9$) and community number ($C = 4, 8$), while keeping the other parameters ($N = 100$, $m = 5$, $\gamma = 3$) fixed. The related communities are also highlighted using different node colours. The figure indicates below each network also the

Normalized Mutual Information (NMI) [16], a measure of performance for evaluating the community detection (see Methods for details), computed by comparing the nPSO ground-truth communities and the ones detected by Louvain [17], which is one of the state-of-the-art community detection algorithms [18]. We notice that the communities are perfectly detected both for $C = 4$ and $C = 8$ at low temperature, suggesting that a meaningful community structure is generated by the proposed model. For the same number of communities, if the temperature is increased the performance slightly decreases, because more inter-community links are established in the network, causing as expected higher rate of wrong assignments by the community detection algorithm. Since these first examples suggested that the networks generated by the nPSO model could be adopted as an interesting benchmark for investigations on community detection, in the next section we compared the performance of different community detection algorithms on nPSO networks.

**Comparison of community detection algorithms on nPSO networks**

We compared the performance of four state-of-the-art approaches [18] (Louvain [17], Infomap [19], Walktrap [20], Label propagation [21]) across networks generated using diverse nPSO parameter combinations.

Fig. 3 reports the mean NMI performance and related standard errors (10 network realizations considered for each parameter combination) of the community detection algorithms applied to nPSO networks with 4 communities. The results indicate that overall Louvain appears as the strongest approach, with an almost perfect detection over different values of network size, average node degree and temperature. Infomap highlights problems in correctly detecting the communities when there are too many inter-community links, as can be seen for $N = 100$ and increasing temperature. The higher temperature in fact leads to a higher number of links between nodes that are far in the disk, which increases the mixing between the communities. The performance is more stable for bigger networks, although in general slightly worse than Louvain. Walktrap results as robust as Louvain to the increase of network temperature, but the NMI is slightly lower for $N = 100$ and $N = 1000$. As last, Label propagation, which is the fastest approach, but the one with lowest accuracy, performs worse than the other methods and presents the same problem as Informap for $N = 100$. This issue has been already pointed out in the study of Yang et al. [18], in which it is shown that for a high mixing of the communities Louvain and Walktrap are more robust, whereas Infomap and Label propagation tend to drop in performance. Hence, the nPSO model here proposed seems to provide a good benchmark to test community detection algorithms on networks generated using a latent geometry model which is based on the hyperbolic space.

Fig. 4 reports the NMI performance on networks with 8 communities. Focusing firstly on the performance on bigger networks ($N$ = [500-1000]), it can be noticed that Louvain and Infomap swap their behaviour, with Infomap going close to the perfect community detection and Louvain slightly decreasing its performance. Walktrap, instead, remains quite robust and slightly improves for $N$ = 1000. Label propagation still remains the most unstable, although surpassing Louvain for very low temperature. In the study of Yang et al. [18] it is shown that, when the mixing of the communities is not high, Louvain can slightly underestimate the number of communities for networks of increasing size, which might be the reason of its reduced performance in large networks ($N$ = [500-1000]) with respect to the other approaches. Focusing now on the small size networks ($N$ = 100), Suppl. Fig. 1 highlights that the methods preserve the same ranking with respect to the case with 4 communities, but they all decrease their performance. The reason is that, being the network small and keeping the average node degree constant, the increase of the communities leads to more inter-community links, making the community structure less detectable.

**Comparison of link prediction algorithms on nPSO networks**

We investigated if the nPSO networks could represent a realistic framework also for testing link prediction algorithms. We compared the performance of three state-of-the-art approaches [22], [23] (CRA [24]–[26] , SPM [27], RA [28]) across both PSO and nPSO networks generated using diverse parameter combinations. The aim of this section is to investigate whether using synthetic networks (generated by PSO or nPSO) is possible to replicate the same link-predictor performance obtained on real complex networks.

In the PSO networks, as reported in Fig. 5, the three methods obtain a comparable precision, with RA performing slightly better than CRA in particular for $N$ = 100, which in turn offers a small improvement with respect to SPM for low temperature $T$ = 0.1. The fact that RA performs slightly better than CRA in wiring-prediction of synthetic networks generated by a uniform model without community is expected, because RA does not account for local-community-organization in the network, and therefore it should adhere better than CRA to the community-free structure of the PSO. In the PSO model a node connects to the other nodes in an 'isotropic' manner, it means that there is not any connection preference on the left or right side of the angular coordinates. Whereas, in the nPSO model a node connects to the other nodes in an 'anisotropic' manner, it means that there is a preference for nodes on the left or right side of the angular coordinates in relation to the direction of localization of the community to which that node belongs. In practice, RA is a weighted version of common neighbours similarity that penalizes each common neighbour for its degree 'isotropically' (in the sense that: taking a common neighbour the penalization is the same for each link that contributes to determine its degree). Instead, CRA penalizes each common neighbour for its degree 'anisotropically' (in the sense that: taking a common neighbour the penalization is not effective for links that - although contribute to determine its degree - are connected to other common neighbours and create a local community). Hence, if our rationale is correct, we expect that CRA should clearly outperform RA not only on nPSO but also on real networks, and this improvement should emerge especially for growing network size, because a large number of nodes favours the generation of non-uniform topology.

When the link-prediction techniques are tested on the nPSO networks, as shown in Fig. 6-7, it can be noticed that the introduction of the communities leads to a different ranking. In fact, while for $N = 100$ (small size networks) the performance of the methods remains overall comparable, for increasing network size and particularly for low temperature CRA overcomes the other two link prediction approaches, regardless of the number of communities adopted (4 or 8). In order to check whether these results resemble a real scenario, we tested the link prediction algorithms also on real networks. Table 1-3 summarize the performance of the approaches for small-size and large-size real networks, both in predicting randomly removed links and in predicting links with time information. For small-size networks SPM obtains the highest mean precision-ranking, followed by CRA and, as last, RA. This result does not approximate well what has been seen in the artificial networks, and the main reason can be that the networks present different characteristics. In particular, most of the small-size real networks do not have a marked power-lawness, as reported in Suppl. Table 1, whereas the PSO and nPSO networks are designed to follow a power-law degree distribution. Looking at the large-size networks, which tend to be scale-free (see Suppl. Table 2), CRA obtains the best mean precision-ranking, reproducing the results reported for increasing network size on the nPSO networks, which, therefore, seems to offer a more realistic framework with respect to the original PSO model.

The plots in Fig. 5-7 report as a reference also the performance that is obtained if the links are predicted ranking them by the hyperbolic distances (HD) or the hyperbolic shortest paths (HSP) between the nodes in the original synthetic network. It can be noticed that in the PSO model the HD performance is slightly lower than CRA for $T = 0.1$ and higher than CRA for $T = [0.3, 0.5]$, whereas in the nPSO model, with the introduction of the communities, the HD performance consistently decreases, since most of the non-existing links within each community are at low HD and therefore are top-ranked. The HSP, instead, provide a quite low precision and always lower than the other methods, which could be expected because the links in the PSO and nPSO networks are established only depending on the HD.

Since the real networks tend to present a community structure, these results obtained on the nPSO suggest that embedding a network in the hyperbolic space and using the ranked HD for predicting the links will not generally lead to high values of precision.

In order to prove this we applied the coalescent embedding techniques [29], a topological-based machine learning class of algorithms that provides a fast and efficient hyperbolic embedding, and the other hyperbolic embedding methods: HyperMap [30], HyperMap-CN [31] and LPCS [32]. Given the coordinates of the embedding, we adopted both the HD-ranking and HSP-ranking in the hyperbolic space for predicting randomly removed links in the small-size real networks, the results are reported in Supplementary Table 4. The maximum average precision offered by the techniques is 0.17, which confirms the expectations independently from the mapping method used. However, if the HSP-ranking is adopted, we notice a general increase of performance with respect to the HD-ranking. From a theoretical standpoint, this result indicates that in real networks the geometry might be even more hyperbolic than the one generated using the PSO/nPSO models. From an applied standpoint, the same result suggests that on real networks combining both the geometrical and the topological information (using the HSP) might help to improve the prediction, and actually this is a confirmation of a result already presented by Cannistraci et al. [33] on prediction of protein interactions by network embedding. At last, we underlines that the small difference in performance between HD and HSP detected in the real networks is better resembled by the nPSO model. In fact, using the PSO model, the difference in performance between HD and HSP is large, and this is markedly in disagreement with the results obtained in real networks.

To conclude, we would like to discuss the fact that in the generative procedure of the PSO and nPSO models the links are not always established between the closest nodes, but with a probability dependent on the hyperbolic distance. Therefore, we argue that the usage of the HD-ranking might be not the best solution for link prediction. Theoretically, the HD distribution should be inferred from the network embedding and, assuming that well approximates the original distribution, it should be exploited for sampling the links from the ranking. However, since the main focus of the article is not link prediction, we procrastinate this investigation to future studies.

**nPSO algorithm efficiency in generating networks with higher clustering**

One of the main drawbacks to use the original algorithmic implementation to establish links adopted by the PSO and GPA models also for the nPSO model, is the lack of efficiency in generating networks with communities characterized by high clustering (low temperature). As reported in Fig. 8-10, the computational time for generating PSO networks is in the order of seconds, whereas for nPSO networks of size $N = 1000$ with low temperature $T = 0.1$, it might take almost one hour ($C = 8$) or up to several hours ($C = 4$), depending on the number of communities.

The main reason is the following. Assuming $T > 0$, at each time step $i$ of the generative procedure the new node $i$ picks a randomly chosen existing node $j < i$ and, given that it is not already connected to it, it connects with probability $p(i,j)$, repeating the procedure until it becomes connected to $m$ nodes. However, it is possible to note from the equation in the Methods section 1 that the connection probability $p(i,j)$ decreases both for decreasing temperature and for increasing hyperbolic distance. Therefore, while generating a network with low temperature and where many nodes are at high hyperbolic distance (for instance: a non-uniform PSO model that displays communities presents hyperbolic distances significantly higher than a classical uniform PSO), most of the connection probabilities to the targets will be low. As a consequence, many iterations will be required before that $m$ connections are successfully established.

Here we propose two different algorithmic implementations, whose details are provided in the Methods sections. Fig. 8-10 shows that both the implementations do not present any issue for generating nPSO networks of low temperature. As highlighted in Suppl. Fig. 12-13, the fastest is implementation 3, whose key idea is to sample the target nodes according to the theoretical probabilities $p(i,j)$, and it only requires 5 minutes to generate large-size nPSO networks of $N = 10000$, regardless of the temperature. In order to prove that modifying the implementation for establishing the links we do not bias the generative procedure toward networks with different properties, we report as Suppl. Fig. 2-11 all the simulations of community detection and link prediction repeated on networks generated using implementations 2 and 3. It can be noticed that the results are almost identical to the ones in the main article. Furthermore, in Suppl. Table 3 we report for each PSO and nPSO parameter combination the average clustering coefficient of the networks generated using the 3 different implementations, confirming that this fundamental property of the model is not biased by the adoption of the algorithmic variants.

# Conclusion

Recent studies presented the hyperbolic disk as an adequate space to describe the latent geometry of real complex networks and the PSO model was introduced to generate random geometric graphs in the hyperbolic space, reproducing strong clustering and a scale-free degree distribution [5]. Coupling the hyperbolic space with the preferential attachment of nodes to this space, the GPA model confers to the networks also a community structure, introducing the idea that different angular regions of the hyperbolic disk can have variable level of attractiveness [15]. However, the GPA model does not allow to indicate in input a desired number of communities, neither to control their size and the mixing between them, which is a clear limitation for real applications. For this reason, we here introduced the nonuniform PSO (nPSO) model, which allows to explicitly fix the number of communities and their size by means of a tailored probability distribution on the angular coordinates, and to tune the mixing property through the network temperature.

The nPSO model has been used as a benchmark for testing state-of-the-art community detection approaches, the evaluations through several parameter combinations highlighted two main points: firstly, the communities are always detected with high accuracy by at least one method; secondly, performance limitations arisen in particular conditions for some community detection methods are in agreement with findings produced in previous studies. These points suggest that the model is able to generate a meaningful community structure, which, together with strong clustering and a scale-free degree distribution, are properties commonly observed in real-world networks.

As second main result, we tested state-of-the-art link prediction algorithms both in real and artificial networks, and we showed that the ranking of the methods in the nPSO model were closer to the ones in the real networks with respect to the PSO model. Furthermore, we highlighted that embedding a network in the hyperbolic space and adopting the HD-ranking for suggesting the links more likely to appear will not lead generally to an efficient prediction, pointing out that the usage of the original network node-HD-distribution needs to be investigated.

At last, since the original algorithm implementation to establish links adopted by the PSO and GPA models is computational expensive for generating networks with communities and high clustering, we proposed other two different variants. The faster of the two implementations significantly reduces the computational time, which is independent from the communities and the clustering.

To conclude, we propose the nPSO model as a valid framework able to efficiently generate realistic networks with a fixed number of communities, which can be adopted, among the many possibilities, also as a reliable benchmark for community detection and link prediction.

## Methods

### 1. PSO model

The Popularity-Similarity-Optimization (PSO) model [5] is a generative network model recently introduced in order to describe how random geometric graphs grow in the hyperbolic space. In this model the networks evolve optimizing a trade-off between node popularity, abstracted by the radial coordinate, and similarity, represented by the angular coordinate distance, and they exhibit many common structural and dynamical characteristics of real networks.

The model has four input parameters:

- $m > 0$, which is equal to half of the average node degree;
- $\beta \in (0, 1]$, defining the exponent $\gamma = 1 + 1/\beta$ of the power-law degree distribution;
- $T \geq 0$, which controls the network clustering; the network clustering is maximized at $T = 0$, it decreases almost linearly for $T = [0,1)$ and it becomes asymptotically zero if $T > 1$;
- $\zeta = \sqrt{-K} > 0$, where $K$ is the curvature of the hyperbolic plane. Since changing $\zeta$ rescales the node radial coordinates and this does not affect the topological properties of networks [5], in the rest of the article we will consider $K = -1$.

Building a network of $N$ nodes on the hyperbolic disk requires the following steps:

(1) Initially the network is empty;

(2) At time $i = 1, 2, \ldots, N$ a new node $i$ appears with radial coordinate $r_i = 2\ln(i)$ and angular coordinate $\theta_i$ uniformly sampled in $[0, 2\pi]$; all the existing nodes $j < i$ increase their radial coordinates according to $r_j(i) = \beta r_j + (1 - \beta) r_i$ in order to simulate popularity fading;

(3) If $T = 0$, the new node connects to the $m$ hyperbolically closest nodes; if $T > 0$, the new node picks a randomly chosen existing node $j < i$ and, given that it is not already connected to it, it connects to it with probability

$$p(i,j) = \frac{1}{1 + \exp\left(\frac{h_{ij} - R_i}{2T}\right)}$$

repeating the procedure until it becomes connected to $m$ nodes.

Note that

$$R_i = r_i - 2\ln\left[\frac{2T(1 - e^{-(1-\beta)\ln(i)})}{\sin(T\pi)\,m(1-\beta)}\right]$$

is the current radius of the hyperbolic disk, and

$$h_{ij} = arccosh(\cosh r_i \cosh r_j - \sinh r_i \sinh r_j \cos\theta_{ij})$$

is the hyperbolic distance between node *i* and node *j*, where

$$\theta_{ij} = \pi - \left|\pi - |\theta_i - \theta_j|\right|$$

is the angle between these nodes.

(4) The growing process stops when N nodes have been introduced.

## 2. GPA model

The GPA model is a variation of the original PSO model that couples the latent hyperbolic network geometry with preferential attachment of nodes to this geometry in order to generate networks with strong clustering, scale-free degree distribution and a non-trivial community structure [15].

The procedure to generate a network of *N* nodes is the same described in the previous section for the PSO model, with the main difference that the angular coordinate $\theta_i$ of the new node *i* is assigned as follows:

(a) Sample $\varphi_1, \dots, \varphi_i$ in $[0, 2\pi]$ uniformly at random. The points $(r_i, \varphi_j)$ for $j = 1 \dots i$ represent candidate positions for the node.

(b) Define for each candidate position $(r_i, \varphi_j)$ the attractiveness $A_i(\varphi_j)$ equal to the number of existing nodes that lie within hyperbolic distance $r_i$ from it.

(c) Set the angular coordinate $\theta_i = \varphi_j$ with probability:

$$\Pi_i(\varphi_j) = \frac{A_i(\varphi_j) + \Lambda}{\sum_{k=1}^{i} A_i(\varphi_k) + \Lambda}$$

Where $\Lambda \geq 0$ is a parameter representing the initial attractiveness.

Note that the GPA model has been presented in the related study with only three input parameters, *m*, $\beta$ and $\Lambda$, with the additional parameters of the PSO model considered in the setting $T = 0$ and $K = -1$.

## 3. Nonuniform PSO (nPSO) model

The nonuniform PSO model is a variation of the PSO model introduced in order to confer to the generated networks an adequate community structure, which is lacking in the original model. Since the connection probabilities are inversely proportional to the hyperbolic distances, a uniform distribution of the nodes over the hyperbolic disk does not create agglomerates of nodes that are concentrated on angular sectors and that are more densely connected between each other than with the rest of the network. A nonuniform distribution, instead, allows to do it by generating heterogeneity in angular node arrangement. In particular, without loss of generality, we will concentrate on the Gaussian mixture distribution, which we consider suitable for describing how to build a nonuniform distributed sample of nodes along the angular coordinates of a hyperbolic disk, with communities that emerge in correspondence of the different Gaussians.

A Gaussian mixture distribution is characterized by the following parameters [46]:

- $C > 0$, which is the number of components, each one representative of a community;
- $\mu_{1...C} \in [0, 2\pi]$, which are the means of every component, representing the central locations of the communities in the angular space;
- $\sigma_{1...C} > 0$, which are the standard deviations of every component, determining how much the communities are spread in the angular space; a low value leads to isolated communities, a high value makes the adjacent communities to overlap;
- $\rho_{1...C}$ ($\sum_i \rho_i = 1$), which are the mixing proportions of every component, determining the relative sizes of the communities.

Given the parameters of the PSO model $(m, \beta, T)$ and the parameters of the Gaussian mixture distribution $(C, \mu_{1...C}, \sigma_{1...C}, \rho_{1...C})$, the procedure to generate a network of $N$ nodes is the same described in the section for the uniform case, with the only difference that the angular coordinates of the nodes are not sampled uniformly but according to the Gaussian mixture distribution. Note that, although the means of the components are located in $[0, 2\pi]$, the sampling of the angular coordinate $\theta$ can fall out of this range. In this case, it has to be shifted within the original range, as follows:

- If $\theta < 0 \rightarrow \theta = 2\pi - mod(|\theta|, 2\pi)$
- If $\theta > 2\pi \rightarrow \theta = mod(\theta, 2\pi)$

Although the parameters of the Gaussian mixture distribution allow for the investigation of disparate scenarios, as a first case of study we focused on the most straightforward setting. For a given number of components $C$, we considered their means equidistantly arranged over the angular space, the same standard deviation and equal mixing proportions:

- $\mu_i = \frac{2\pi}{C} * (i - 1) \quad i = 1 ... C$
- $\sigma_1 = \sigma_2 = ... = \sigma_C = \sigma$

- $\rho_1 = \rho_2 = \ldots = \rho_C = \frac{1}{C}$

In particular, in our simulations we fixed the standard deviation to 1/6 of the distance between two adjacent means $\left(\sigma = \frac{1}{6} * \frac{2\pi}{C}\right)$, which allowed for a reasonable isolation of the communities. The community memberships are assigned considering for each node the component whose mean is the closest in the angular space.

### 4. Implementations for link generation

In the Methods section 1, at step (3) of the generative procedure, it is presented how the new node establishes connections to *m* of the existing nodes. In particular, if $T > 0$, the new node $i$ picks a randomly chosen existing node $j < i$ and, given that it is not already connected to it, it connects with probability $p(i,j)$, repeating the procedure until it becomes connected to *m* nodes. At the implementation level, the basic solution in MATLAB code would be:

```
targets = 1:(t-1);
c = 0;
while c < m
      rand_p = rand(1);
      idx = randi(length(targets));
      j = targets(idx);
      if p(j) > rand_p
           x(t,j) = 1;
           c = c + 1;
           targets(idx) = [];
      end
end
```

Where *t* is the new node, *p* is the vector of connection probabilities from *t* to the previous nodes and *x* is the adjacency matrix of the network. We will refer to this as implementation 1.

As commented in the main article, this implementation have issues of time performance in specific cases. In fact, it is possible to note from the equation in the Methods section 1 that the connection probability $p(i,j)$ decreases both for decreasing temperature and for increasing hyperbolic distance. Therefore, while generating a network with low temperature and where many nodes are at high hyperbolic distance (for example sampling the angular coordinates from a Gaussian mixture distribution with 4 communities), most of the connection probabilities to the targets will be low. As a consequence, the *if* statement will result *false* in many iterations and the *while* loop will require a relevant computational time before that *m* connections are successfully established.

In order to solve this issue, we note that at each loop iteration the connection probabilities to the target nodes (excluded the ones already connected) do not always cover the full range [0,

1]. In particular, at each iteration the maximum of these probabilities $p_{max} = \max\limits_{j \in \{targets\}} p(i,j)$ will be usually lower than 1. Since it is known a priori that any random sampling $rand\_p > p_{max}$ will necessarily bring to a rejection of the connection independently from the target node chosen, the sampling range of the random probability $rand\_p$ can be restricted to $[0, p_{max}]$. In the critical conditions previously mentioned, where most of the connection probabilities are low, this adjustment can bring to a considerable speedup without biasing the link generation procedure.

Furthermore, using a programming language optimized for vector operations (i.e. MATLAB) rather than loop-based code, a further improvement can be done. Since vector operations are faster than loop-based operations, at each iteration *m* attempts of connection to target nodes can be done at once, reducing the number of iterations required to successfully establish *m* connections. The MATLAB code of the implementation would be:

```
targets = 1:(t-1);
c = 0;
while c < m
    if length(targets) > m
        idx = randsample(length(targets),m);
    else
        idx = 1:length(targets);
    end
    rand_p = rand(1,length(idx)) * max(p(targets));
    idx = idx(p(targets(idx)) > rand_p);
    if length(idx) > m - c
        idx = randsample(idx,m - c);
    end
    x(t,targets(idx)) = 1;
    targets(idx) = [];
    c = c + length(idx);
end
```

Where *t* is the new node, *p* is the vector of connection probabilities from *t* to the previous nodes and *x* is the adjacency matrix of the network. We will refer to this as implementation 2.

Note that, while the sampling of *m* targets at each iteration is an adjustment convenient only when using a programming language optimized for vectorization, the restriction of the probability sampling to the range $[0, p_{max}]$ is valid in general.

A further variant that we propose is to sample the target nodes according to the theoretical probabilities $p(i,j)$. This solution ensures that at every iteration new connections are successfully established, making the procedure faster. The MATLAB code of the implementation would be:

```
targets = 1:(t-1);
c = 0;
```

```
while c < m
    idx = unique(randsample(length(targets),m,1,p(targets)));
    if length(idx) > m - c
        idx = randsample(idx,m - c);
    end
    x(t,targets(idx)) = 1;
    targets(idx) = [];
    c = c + length(idx);
end
```

Where *t* is the new node, *p* is the vector of connection probabilities from *t* to the previous nodes and *x* is the adjacency matrix of the network. We will refer to this as implementation 3.

Note that, as for the previous implementation, the sampling of *m* targets at each iteration is an adjustment convenient only when using a programming language optimized for vectorization, whereas the idea of sampling according to the theoretical probabilities is valid in general.

The computational speed of the three implementations as well as the equivalence of the networks generated is discussed in the main text.

**Hardware and software**

MATLAB code has been used for all the simulations, carried out partly on a workstation under Windows 8.1 Pro with 512 GB of RAM and 2 Intel(R) Xenon(R) CPU E5-2687W v3 processors with 3.10 GHz, and partly in the ZIH-Cluster Taurus of the TU Dresden.

**Funding**

Work in the CVC laboratory was supported by the independent research group leader starting grant of the Technische Universität Dresden. AM was partially supported by the funding provided by the Free State of Saxony in accordance with the Saxon Scholarship Program Regulation, awarded by the Studentenwerk Dresden based on the recommendation of the board of the Graduate Academy of TU Dresden.

**Acknowledgements**

We thank Alexander Mestiashvili and the BIOTEC System Administrators for their IT support, Claudia Matthes for the administrative assistance and the Centre for Information Services and High Performance Computing (ZIH) of the TUD.

**Author contributions**

CVC invented the nonuniform PSO model and designed the numerical experiments. AM implemented the code and performed the computational analysis. Both the authors analysed and interpreted the results. AM wrote the draft of the article according to CVC suggestions and CVC corrected and improved it to arrive to the final draft. CVC designed the figures and AM realized them. CVC planned, directed and supervised the study.

**Competing interests**

The authors declare no competing financial interests.

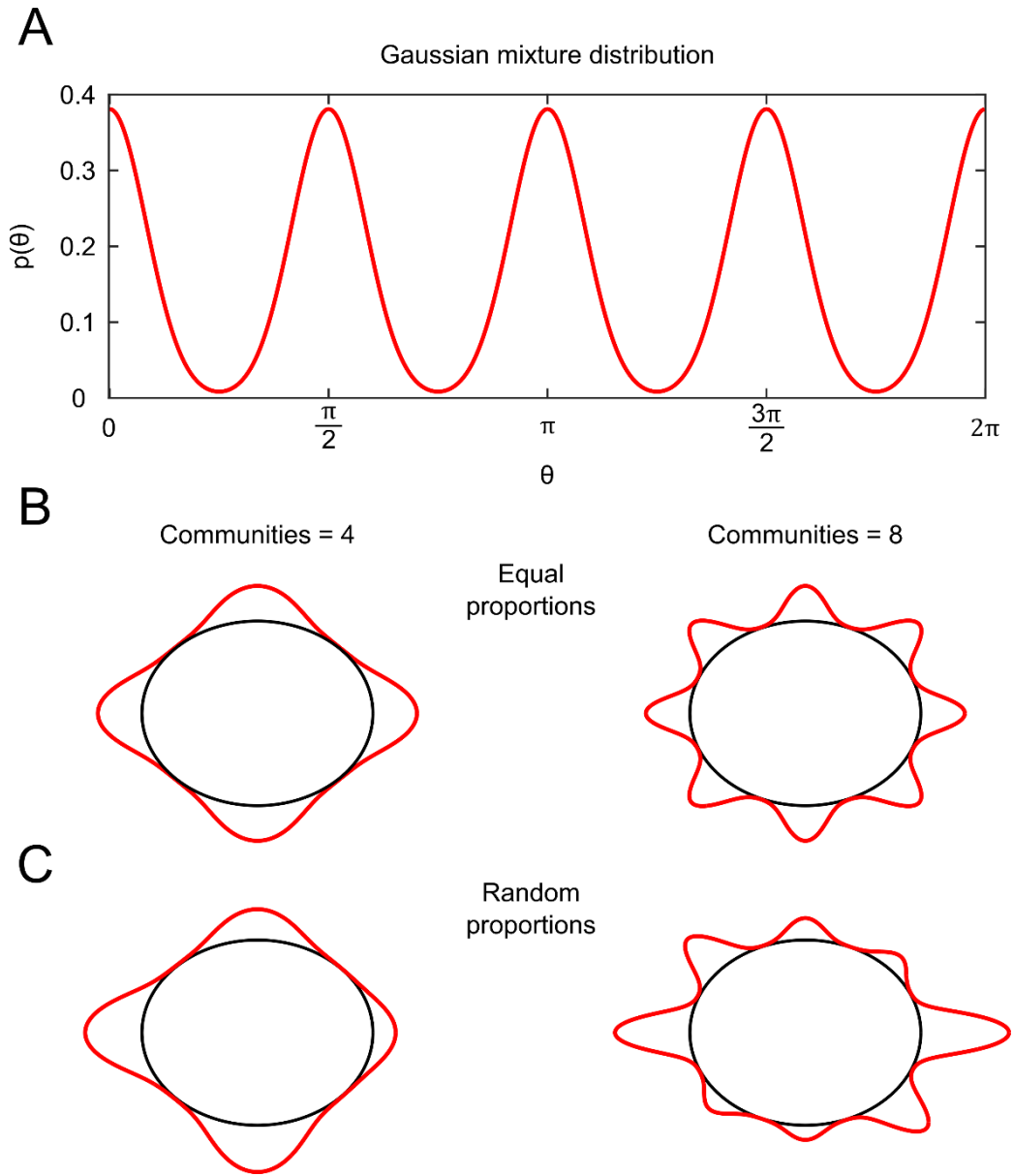

**Figure 1. Nonuniform distribution of angular coordinates.**
The figure shows examples of nonuniform distributions used for sampling the angular coordinates of the nodes. The distributions are generated using a Gaussian mixture model, with as many components as the number of the desired communities, placing the mean of the components equidistantly over the angular space and with equal standard deviations, whose value is chosen as 1/6 of the distance between two adjacent means. (A) Plot of the Gaussian mixture distribution using 4 components having the same mixing proportion. (B) Representation of the Gaussian mixture distribution, using 4 and 8 components having the same mixing proportion, along the angular space of the hyperbolic disk. (C) Representation of the Gaussian mixture distribution, using 4 and 8 components having the random mixing proportion, along the angular space of the hyperbolic disk.

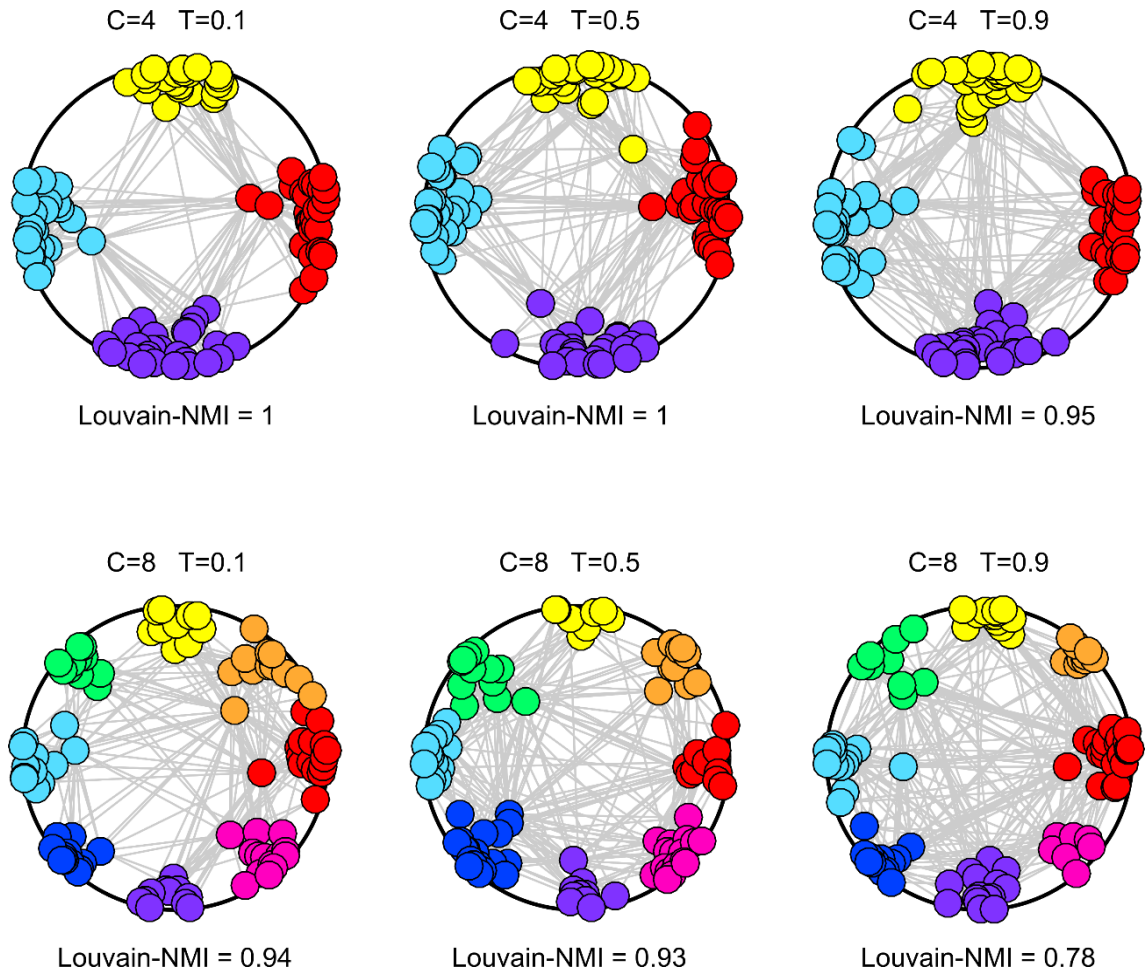

**Figure 2. Communities generated using the nPSO model.**
Synthetic networks have been generated using the nPSO model with parameters $\gamma = 3$ (power-law degree distribution exponent), $m = 5$ (half of average degree), $T = [0.1, 0.5, 0.9]$ (temperature, inversely related to the clustering coefficient), $N = 100$ (network size) and C = [4, 8] (communities). For each combination of parameters, 10 networks have been generated. For each network the Louvain community detection method has been executed and the communities detected have been compared to the annotated ones computing the Normalized Mutual Information (NMI). The plots show for each parameter combination a representation in the hyperbolic space of the network that obtained the highest NMI, whose value is reported. The nodes are coloured according to the communities as generated by the nPSO model.

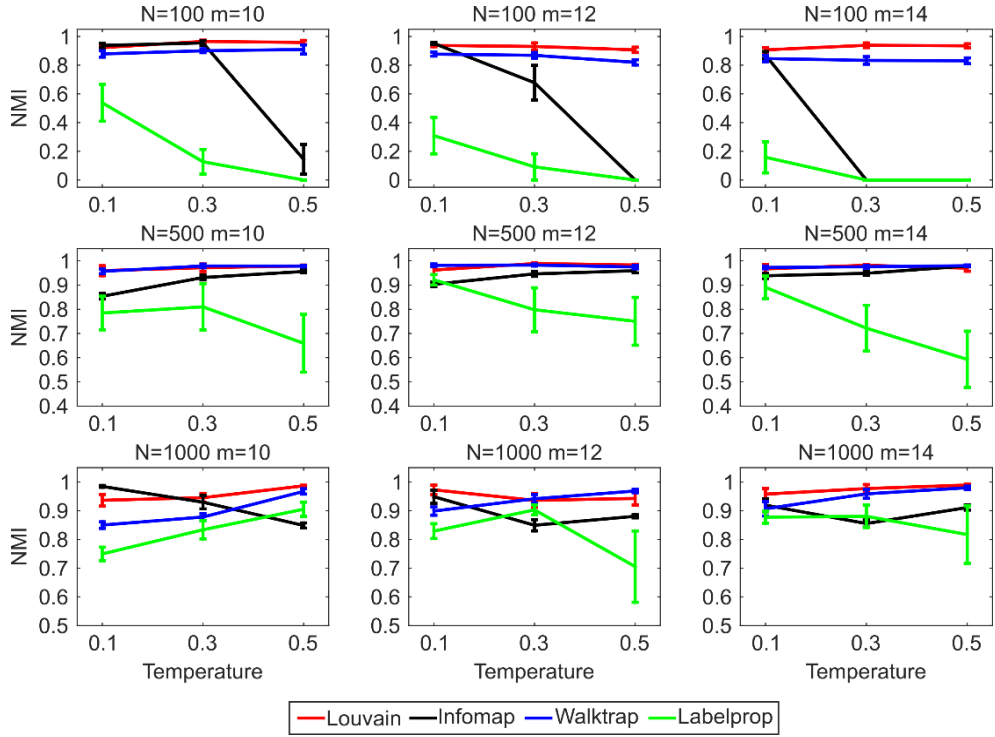

**Figure 3. Community detection on nPSO networks with 4 communities.**
Synthetic networks have been generated using the nPSO model with parameters $\gamma = 3$ (power-law degree distribution exponent), $m = [10, 12, 14]$ (half of average degree), $T = [0.1, 0.3, 0.5]$ (temperature, inversely related to the clustering coefficient), $N = [100, 500, 1000]$ (network size) and 4 communities. The values chosen for the parameter $m$ are centered around the average $m$ computed on the dataset of small-size real networks. The values chosen for $N$ and $T$ are intended to cover the range of network size and clustering coefficient observed in the dataset of small-size real networks. Since the average $\gamma$ estimated on the dataset of small-size real networks is higher than the typical range $2 < \gamma < 3$, we choose $\gamma = 3$. For each combination of parameters, 10 networks have been generated. For each network the community detection methods Louvain, Infomap, Walktrap and Label propagation have been executed and the communities detected have been compared to the annotated ones computing the Normalized Mutual Information (NMI). The plots report for each parameter combination the mean NMI and standard error over the random iterations. The networks have been generated using the implementation 1, the same simulations using implementations 2 and 3 are reported in Supplementary Information.

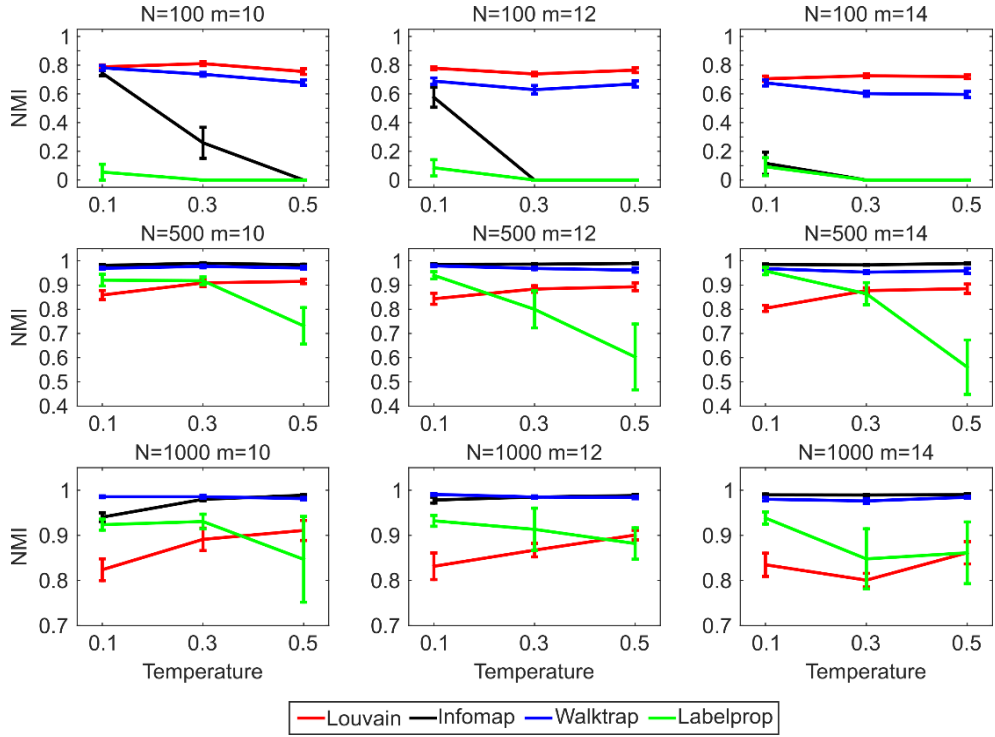

**Figure 4. Community detection on nPSO networks with 8 communities.**
Synthetic networks have been generated using the nPSO model with parameters $\gamma = 3$ (power-law degree distribution exponent), $m = [10, 12, 14]$ (half of average degree), $T = [0.1, 0.3, 0.5]$ (temperature, inversely related to the clustering coefficient), $N = [100, 500, 1000]$ (network size) and 8 communities. The values chosen for the parameter $m$ are centered around the average $m$ computed on the dataset of small-size real networks. The values chosen for $N$ and $T$ are intended to cover the range of network size and clustering coefficient observed in the dataset of small-size real networks. Since the average $\gamma$ estimated on the dataset of small-size real networks is higher than the typical range $2 < \gamma < 3$, we choose $\gamma = 3$. For each combination of parameters, 10 networks have been generated. For each network the community detection methods Louvain, Infomap, Walktrap and Label propagation have been executed and the communities detected have been compared to the annotated ones computing the Normalized Mutual Information (NMI). The plots report for each parameter combination the mean NMI and standard error over the random iterations. The networks have been generated using the implementation 1, the same simulations using implementations 2 and 3 are reported in Supplementary Information.

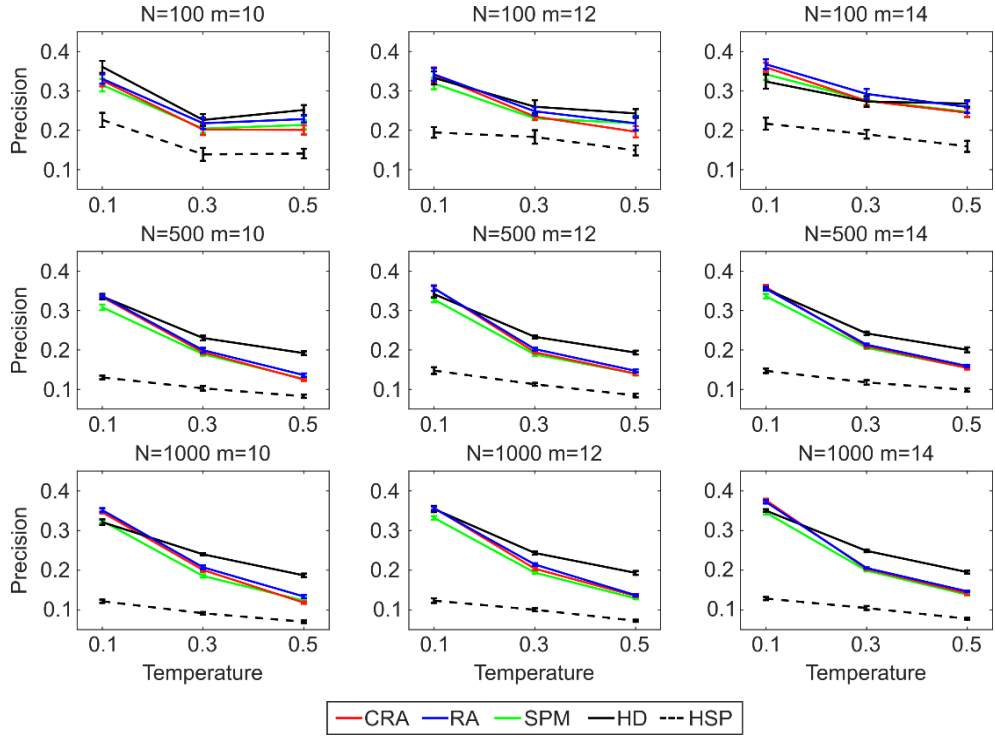

**Figure 5. Link prediction on PSO networks.**
Synthetic networks have been generated using the PSO model with parameters $\gamma = 3$ (power-law degree distribution exponent), $m = [10, 12, 14]$ (half of average degree), $T = [0.1, 0.3, 0.5]$ (temperature, inversely related to the clustering coefficient) and $N = [100, 500, 1000]$ (network size). The values chosen for the parameter $m$ are centered around the average $m$ computed on the dataset of small-size real networks. The values chosen for $N$ and $T$ are intended to cover the range of network size and clustering coefficient observed in the dataset of small-size real networks. Since the average $\gamma$ estimated on the dataset of small-size real networks is higher than the typical range $2 < \gamma < 3$, we choose $\gamma = 3$. For each combination of parameters, 10 networks have been generated. For each network 10% of links have been randomly removed and the algorithms have been executed in order to assign likelihood scores to the non-observed links in these reduced networks. In order to evaluate the performance, the links are ranked by likelihood scores and the precision is computed as the percentage of removed links among the top-$r$ in the ranking, where $r$ is the total number of links removed. The plots report for each parameter combination the mean precision and standard error over the random iterations. The networks have been generated using the implementation 1, the same simulations using implementations 2 and 3 are reported in Supplementary Information.

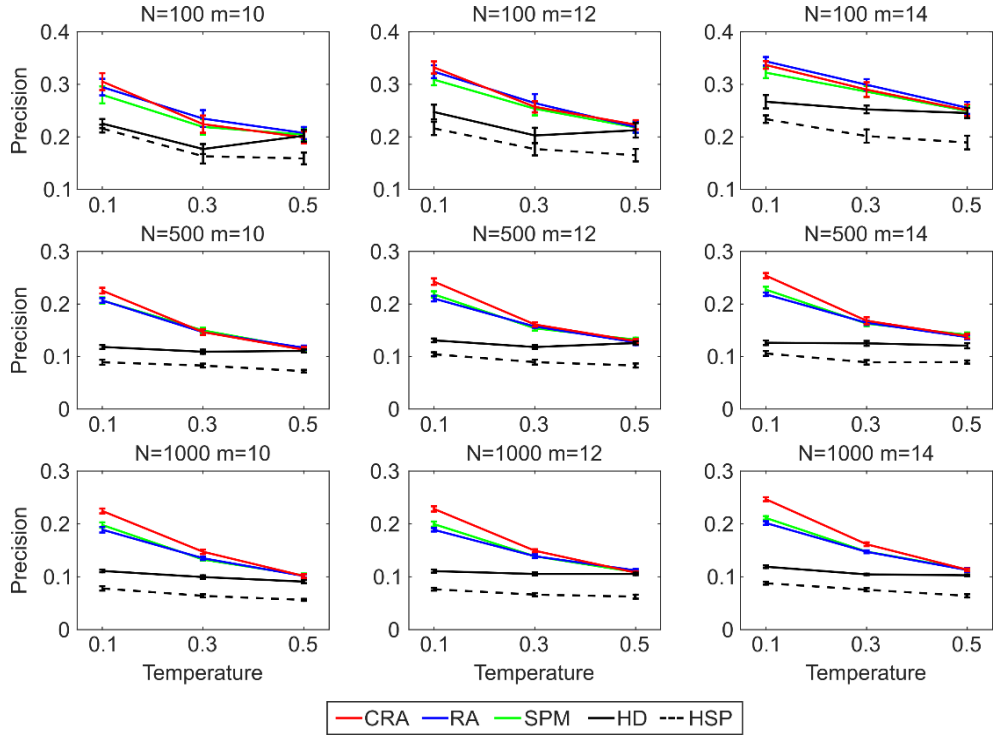

**Figure 6. Link prediction on nPSO networks with 4 communities.**
Synthetic networks have been generated using the nPSO model with parameters $\gamma = 3$ (power-law degree distribution exponent), $m = [10, 12, 14]$ (half of average degree), $T = [0.1, 0.3, 0.5]$ (temperature, inversely related to the clustering coefficient), $N = [100, 500, 1000]$ (network size) and 4 communities. The values chosen for the parameter $m$ are centered around the average $m$ computed on the dataset of small-size real networks. The values chosen for $N$ and $T$ are intended to cover the range of network size and clustering coefficient observed in the dataset of small-size real networks. Since the average $\gamma$ estimated on the dataset of small-size real networks is higher than the typical range $2 < \gamma < 3$, we choose $\gamma = 3$. For each combination of parameters, 10 networks have been generated. For each network 10% of links have been randomly removed and the algorithms have been executed in order to assign likelihood scores to the non-observed links in these reduced networks. In order to evaluate the performance, the links are ranked by likelihood scores and the precision is computed as the percentage of removed links among the top-$r$ in the ranking, where $r$ is the total number of links removed. The plots report for each parameter combination the mean precision and standard error over the random iterations. The networks have been generated using the implementation 1, the same simulations using implementations 2 and 3 are reported in Supplementary Information.

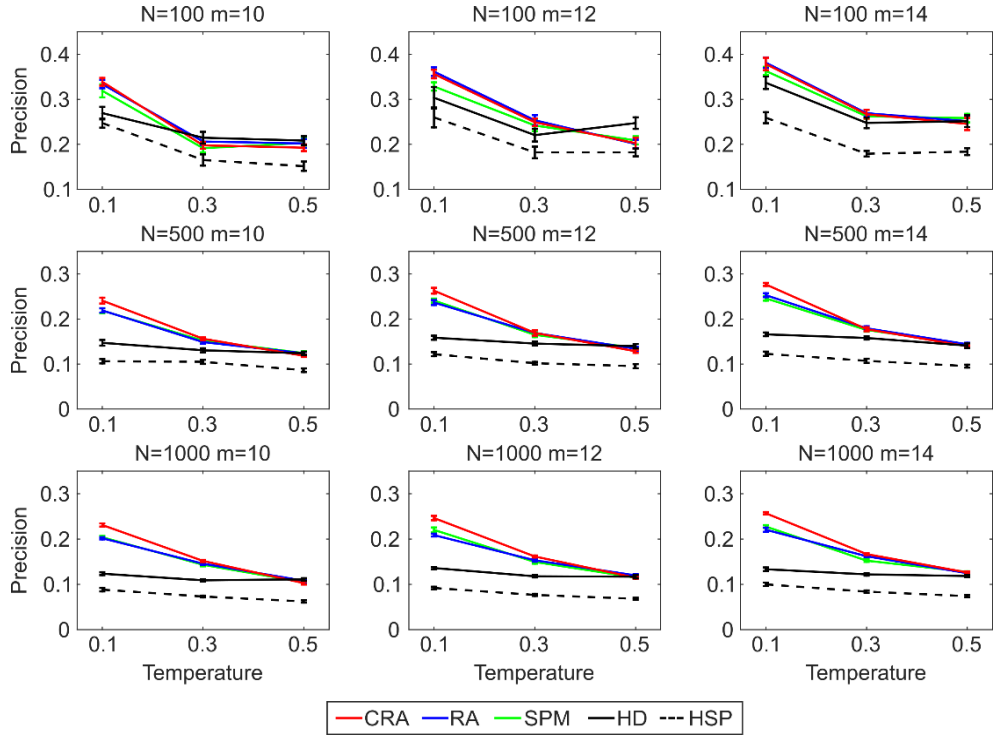

**Figure 7. Link prediction on nPSO networks with 8 communities.**
Synthetic networks have been generated using the nPSO model with parameters $\gamma = 3$ (power-law degree distribution exponent), $m = [10, 12, 14]$ (half of average degree), $T = [0.1, 0.3, 0.5]$ (temperature, inversely related to the clustering coefficient), $N = [100, 500, 1000]$ (network size) and 8 communities. The values chosen for the parameter $m$ are centered around the average $m$ computed on the dataset of small-size real networks. The values chosen for $N$ and $T$ are intended to cover the range of network size and clustering coefficient observed in the dataset of small-size real networks. Since the average $\gamma$ estimated on the dataset of small-size real networks is higher than the typical range $2 < \gamma < 3$, we choose $\gamma = 3$. For each combination of parameters, 10 networks have been generated. For each network 10% of links have been randomly removed and the algorithms have been executed in order to assign likelihood scores to the non-observed links in these reduced networks. In order to evaluate the performance, the links are ranked by likelihood scores and the precision is computed as the percentage of removed links among the top-$r$ in the ranking, where $r$ is the total number of links removed. The plots report for each parameter combination the mean precision and standard error over the random iterations. The networks have been generated using the implementation 1, the same simulations using implementations 2 and 3 are reported in Supplementary Information.

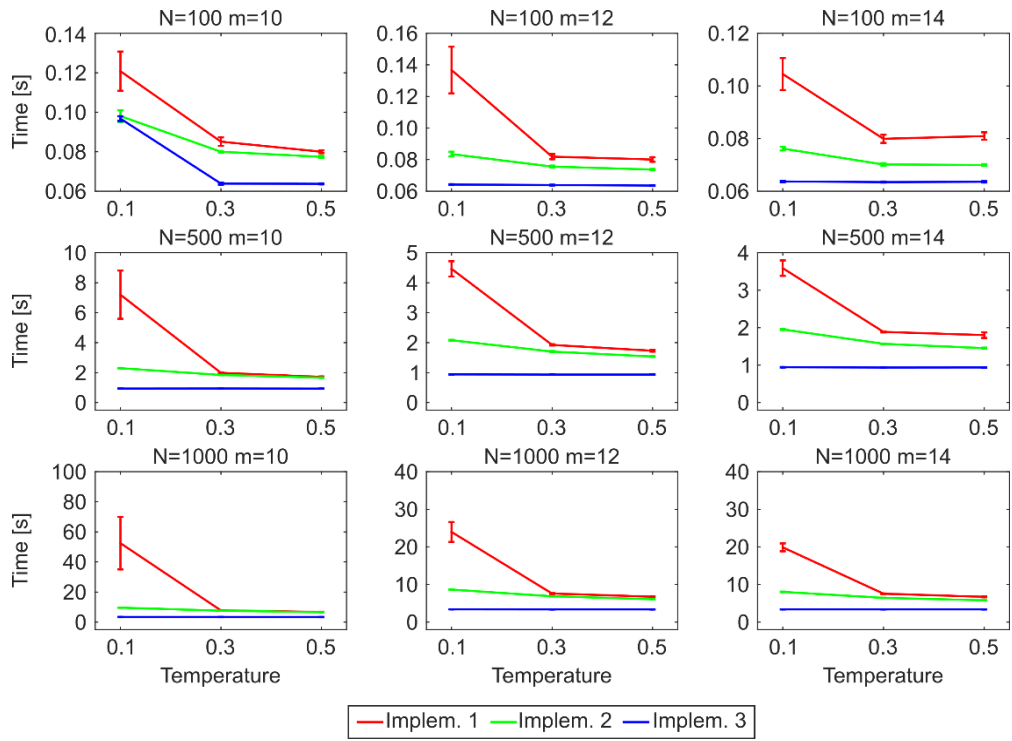

**Figure 8. Time performance for generating PSO networks.**
Synthetic networks have been generated using the PSO model with parameters $\gamma = 3$ (power-law degree distribution exponent), $m = [10, 12, 14]$ (half of average degree), $T = [0.1, 0.3, 0.5]$ (temperature, inversely related to the clustering coefficient) and $N = [100, 500, 1000]$ (network size). For each combination of parameters, 10 networks have been generated using the 3 different implementations. The plots report for each parameter combination the mean computational time and standard error over the random iterations.

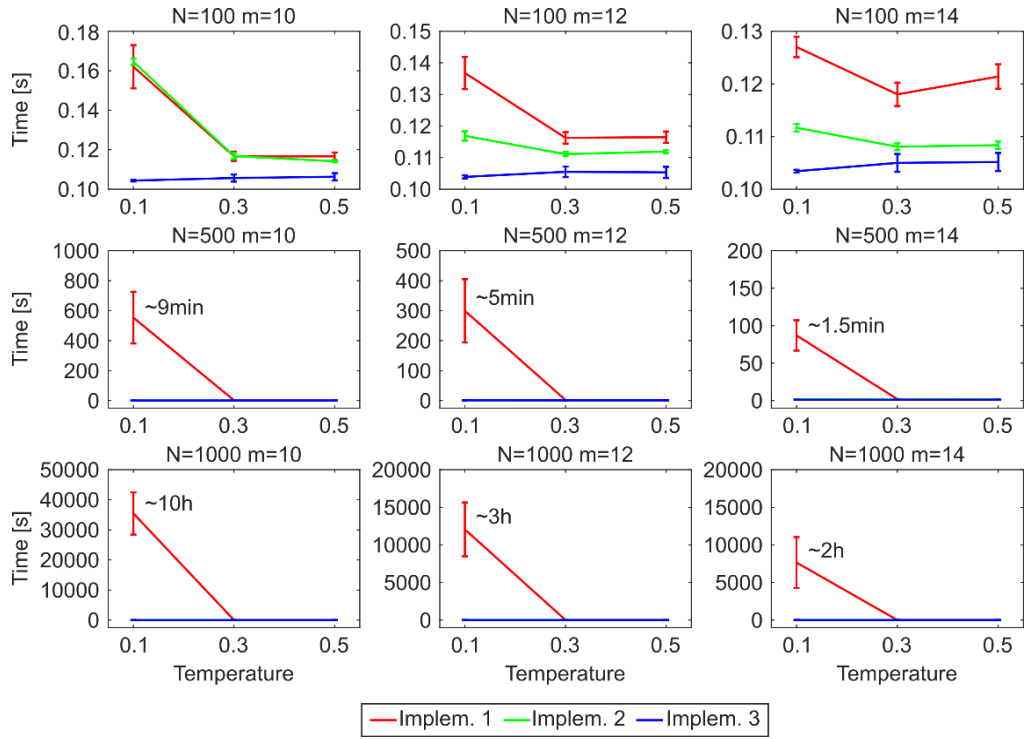

**Figure 9. Time performance for generating PSO networks with 4 communities.**
Synthetic networks have been generated using the PSO model with parameters $\gamma = 3$ (power-law degree distribution exponent), $m = [10, 12, 14]$ (half of average degree), $T = [0.1, 0.3, 0.5]$ (temperature, inversely related to the clustering coefficient), $N = [100, 500, 1000]$ (network size) and 4 communities. For each combination of parameters, 10 networks have been generated using the 3 different implementations. The plots report for each parameter combination the mean computational time and standard error over the random iterations. Due to the different scale of implementation 1, a more detailed comparison of implementations 2 and 3 is provided in Supplementary Information.

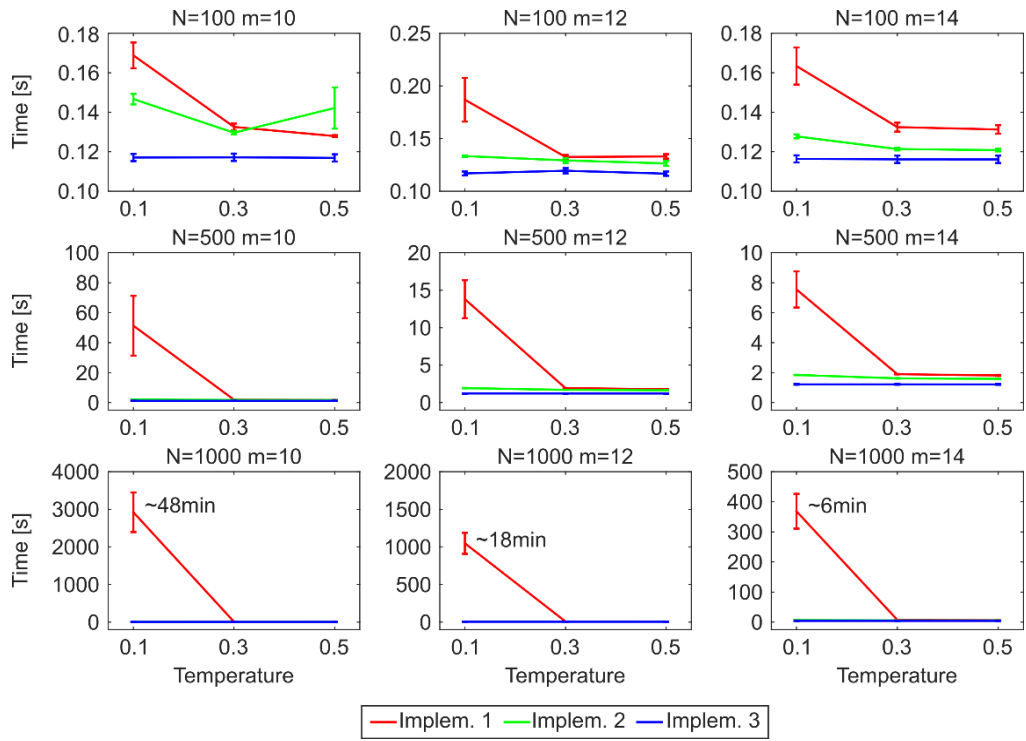

**Figure 10. Time performance for generating nPSO networks with 8 communities.**
Synthetic networks have been generated using the nPSO model with parameters $\gamma = 3$ (power-law degree distribution exponent), $m = [10, 12, 14]$ (half of average degree), $T = [0.1, 0.3, 0.5]$ (temperature, inversely related to the clustering coefficient), $N = [100, 500, 1000]$ (network size) and 8 communities. For each combination of parameters, 10 networks have been generated using the 3 different implementations. The plots report for each parameter combination the mean computational time and standard error over the random iterations. Due to the different scale of implementation 1, a more detailed comparison of implementations 2 and 3 is provided in Supplementary Information.

**Table 1. Link prediction on small-size real networks.**
For each network 10% of links have been randomly removed (100 iterations) and the algorithms have been executed in order to assign likelihood scores to the non-observed links in these reduced networks. In order to evaluate the performance, the links are ranked by likelihood scores and the precision is computed as the percentage of removed links among the top-$r$ in the ranking, where $r$ is the total number of links removed. The table reports for each network the mean precision over the random iterations. The last rows show the mean precision and the mean ranking over the entire dataset. For each network the best method is highlighted in bold. The networks are sorted by increasing number of nodes $N$.

|  | SPM | CRA | RA |
| --- | --- | --- | --- |
| mouse neural | 0.02 | **0.11** | 0.07 |
| karate | 0.17 | **0.20** | 0.14 |
| dolphins | 0.13 | **0.14** | 0.10 |
| macaque neural | **0.72** | 0.56 | 0.51 |
| polbooks | 0.17 | 0.17 | **0.21** |
| ACM2009 contacts | 0.26 | **0.27** | **0.27** |
| football | 0.31 | **0.36** | 0.27 |
| physicians innovation | 0.07 | 0.07 | **0.08** |
| manufacturing email | **0.51** | 0.42 | 0.43 |
| littlerock foodweb | **0.84** | 0.15 | 0.10 |
| jazz | **0.65** | 0.56 | 0.54 |
| residence hall friends | **0.28** | 0.24 | 0.25 |
| haggle contacts | **0.62** | 0.57 | 0.58 |
| worm nervoussys | **0.16** | 0.12 | 0.10 |
| netsci | 0.41 | 0.50 | **0.59** |
| infectious contacts | **0.37** | 0.34 | 0.35 |
| flightmap | **0.75** | 0.54 | 0.56 |
| email | **0.16** | **0.16** | 0.15 |
| polblog | **0.23** | 0.17 | 0.13 |
| mean precision | **0.36** | 0.30 | 0.29 |
| mean ranking | **1.66** | 2.05 | 2.29 |

**Table 2. Link prediction on large-size real networks.**
For each network 10% of links have been randomly removed (10 iterations) and the algorithms have been executed in order to assign likelihood scores to the non-observed links in these reduced networks. In order to evaluate the performance, the links are ranked by likelihood scores and the precision is computed as the percentage of removed links among the top-$r$ in the ranking, where $r$ is the total number of links removed. The table reports for each network the mean precision over the random iterations. The last rows show the mean precision and the mean ranking over the entire dataset. For each network the best method is highlighted in bold. The networks are sorted by increasing number of nodes $N$.

|                | CRA      | SPM      | RA      |
|----------------|----------|----------|---------|
| odlis          | **0.12** | 0.08     | 0.10    |
| advogato       | **0.16** | 0.15     | 0.14    |
| arxiv astroph  | 0.53     | **0.67** | 0.64    |
| thesaurus      | **0.08** | 0.07     | 0.03    |
| arxiv hepth    | 0.22     | **0.27** | 0.20    |
| ARK201012      | **0.16** | 0.11     | 0.16    |
| facebook       | **0.11** | 0.10     | 0.06    |
| mean precision | 0.20     | **0.21** | 0.19    |
| mean ranking   | **1.5**  | 2        | 2.5     |

**Table 3. Link prediction in time on AS Internet networks.**
Six AS Internet network snapshots are available from September 2009 to December 2010, at time steps of 3 months. For every snapshot at times $i = [1, 5]$ the algorithms have been executed in order to assign likelihood scores to the non-observed links and the link prediction performance has been evaluated computing the precision with respect to every future time point $j = [i+1, 6]$. Considering a pair of time points $(i, j)$, the non-observed links at time $i$ are ranked by decreasing likelihood scores and the precision is computed as the percentage of links that appear at time $j$ among the top-$r$ in the ranking, where $r$ is the total number of non-observed links at time $i$ that appear at time $j$. Non-observed links at time $i$ involving nodes that disappear at time $j$ are not considered in the ranking. The table reports for each method a 5-dimensional upper triangular matrix, containing as element $(i, j)$ the precision of the link prediction from time $i$ to time $j+1$. On the bottom-right side, the methods are ranked by the mean precision computed over all the time combinations, the mean ranking is also shown. For each comparison the best method is highlighted in bold.

| | | CRA | | | | | RA | | |
|---|---|---|---|---|---|---|---|---|---|
| **0.11** | **0.12** | **0.13** | **0.14** | **0.14** | 0.10 | 0.11 | 0.11 | 0.12 | 0.12 |
| | **0.12** | **0.13** | **0.14** | **0.14** | | 0.09 | 0.10 | 0.11 | 0.12 |
| | | **0.12** | **0.13** | **0.14** | | | 0.09 | 0.11 | 0.12 |
| | | | **0.12** | **0.13** | | | | 0.10 | 0.11 |
| | | | | **0.12** | | | | | 0.10 |

| | | SPM | | | | mean precision | mean ranking |
|---|---|---|---|---|---|---|---|
| 0.08 | 0.09 | 0.09 | 0.10 | 0.11 | **CRA** | **0.13** | **1** |
| | 0.07 | 0.08 | 0.09 | 0.10 | RA | 0.11 | 2 |
| | | 0.08 | 0.09 | 0.10 | SPM | 0.09 | 3 |
| | | | 0.08 | 0.09 | | | |
| | | | | 0.09 | | | |

# Supplementary Information

## 1. Link prediction methods

### 1.1. Cannistraci-Resource-Allocation (CRA)

Cannistraci-Resource-Allocation (CRA) is a local-based, parameter-free and model-based deterministic rule for topological link prediction in both monopartite [1] and bipartite networks [2], [3]. It is based on the *local-community-paradigm* (LCP) which is a bioinspired theory recently proposed in order to model local-topology-dependent link-growth in a class of real complex networks characterized by the development of diverse, overlapping and hierarchically organized local-communities [1]. Being a local-community-based method, it assigns to every candidate interaction a likelihood score looking only at the neighbours nodes and their cross-interactions. In particular, the paradigmatic shift introduced by the LCP is to consider not only the common neighbours of the interacting nodes but also the links between those common neighbours, which in practice form all together a local community (Fig. 1). For each candidate interaction between nodes $i$ and $j$, the score is assigned according to the following equation [1]:

$$CRA(i,j) = \sum_{k \in \Phi(i) \cap \Phi(j)} \frac{|\varphi(k)|}{|\Phi(k)|}$$

Where:

$k$: common neighbour of nodes $i$ and $j$

$\Phi(i)$: set of neighbours of node $i$

$|\Phi(k)|$: cardinality of set $\Phi(k)$, equivalent to the degree of $k$

$\varphi(k)$: sub-set of neighbours of $k$ that are also common neighbours of $i$ and $j$

$|\varphi(k)|$: equivalent to the local community degree of $k$ (see Fig.1)

The higher the CRA score, the higher the likelihood that the interaction exists, therefore the candidate interactions are ranked by decreasing CRA scores and the obtained ranking is the link prediction result. The method has been implemented in MATLAB. The code is available at: https://sites.google.com/site/carlovittoriocannistraci/

### 1.2. Resource-Allocation (RA)

Resource-Allocation (RA) is a local-based, parameter-free and model-based deterministic rule for topological link prediction [4], motivated by the resource allocation process taking place in networks. Considering a pair of nodes that are not directly connected, one node can send some resource to the other one through their common neighbours, which play the role of transmitters. It assumes the simplest case where every transmitter equally distributes a unit of resource between its neighbours. For each candidate interaction between nodes $i$ and $j$, the score is assigned according to the following equation [4]:

$$RA(i,j) = \sum_{k \in \Phi(i) \cap \Phi(j)} \frac{1}{|\Phi(k)|}$$

Where:

$k$: common neighbour of nodes $i$ and $j$

$\Phi(i)$: set of neighbours of node $i$

$|\Phi(k)|$: cardinality of set $\Phi(k)$, equivalent to the degree of $k$

The higher the RA score, the higher the likelihood that the interaction exists, therefore the candidate interactions are ranked by decreasing RA scores and the obtained ranking is the link prediction result. The method has been implemented in MATLAB.

### 1.3. Structural Perturbation Method (SPM)

SPM is a structural perturbation method that relies on a theory similar to the first-order perturbation in quantum mechanics [5]. Unlike CRA and RA, it is a global approach, meaning that it exploits the information of the complete adjacency matrix in order to compute the likelihood score to assign to every candidate interaction. A high-level description of the procedure is the following:

1) Randomly remove a subset of the edges $\Delta E$ (usually 10%) from the network adjacency matrix $x$, obtaining a reduced adjacency matrix $x^R$.
2) Compute the eigenvalues and eigenvectors of $x^R$.
3) Considering $\Delta E$ as a perturbation of $x^R$, construct the perturbed matrix $\tilde{x}$ via a first-order approximation that allows the eigenvalues to change while keeping fixed the eigenvectors.
4) Repeat steps 1-3 for 10 independent iterations and take the average of the perturbed matrices $\tilde{x}$.

The idea behind the method is that a missing part of the network is predictable if it does not significantly change the structural features of the observable part, represented by the eigenvectors of the matrix. If this is the case, the perturbed matrices should be good approximations of the original network [5]. The entries of the average perturbed matrix represents the scores for the candidate links. The higher the score the greater the likelihood that the interaction exists, therefore the candidate interactions are ranked by decreasing scores and the obtained ranking represents the link prediction result.

The MATLAB implementation of the method has been provided by the authors.

**1.4. Coalescent embedding**

The expression coalescent embedding refers to a topological-based machine learning class of algorithms that exploits nonlinear unsupervised dimensionality reduction to infer the nodes angular coordinates in the hyperbolic space [6]. The techniques are able to perform a fast and accurate mapping of a network in the 2D hyperbolic disk, the 3D hyperbolic sphere, and potentially also in higher dimensions.

The first step of the algorithm for a 2D embedding consists in weighting the network in order to suggest geometrical distances between connected nodes, since it has been shown that improves the mapping accuracy [6]. If the network is unweighted, the topological-based pre-weighting rules repulsion-attraction (RA) or edge-betweenness-centrality (EBC) can be applied. The rules are devised to suggest geometrical distances between the connected nodes, using either local (RA) or global (EBC) topological information [6].

Given the weighted network, the second step consists in performing the nonlinear dimensionality reduction. Two different kinds of machine learning approaches can be used, manifold-based (LE, ISO, ncISO) or Minimum-Curvilinearity-based (MCE, ncMCE). The details about which dimensions of the embedding should be considered are provided in the original publication [6].

In order to assign the angular coordinates in the 2D embedding space, either a circular adjustment or an equidistant angular adjustment (EA) can be considered. The circular adjustment for the manifold-based approaches consists in exploiting directly the polar coordinates of the 2D reduced space, whereas for the Minimum-Curvilinearity-based in rearranging the node points on the circumference following the same ordering of the 1D reduced space and proportionally preserving the distances. Using the equidistant angular adjustment, instead, the nodes are equidistantly arranged on the circumference, which might help to correct for short-range angular noise present in the embedding [6].

The radial coordinates are assigned according to a mathematical formula which takes into account both the position of the nodes in their ranking by degree and the scale-freeness of the node degree distribution [7]. The exponent $\gamma$ of the power-law degree distribution has been fitted using the MATLAB script *plfit.m*, a procedure described by Clauset et al. [8] and released at http://www.santafe.edu/~aaronc/powerlaws/.

In the link prediction application, the hyperbolic distances (HD) are computed between the nodes in the hyperbolic space and the candidate interactions are ranked by increasing HD, the obtained ranking is the link prediction result.

In a second variant that combines both geometrical and topological information, the network is weighted using the HD and the hyperbolic shortest paths (HSP) are computed as sum of the HD over the shortest path between each pair of nodes. The candidate interactions are ranked by increasing HSP and the obtained ranking is the link prediction result.

The method has been implemented in MATLAB.

### 1.5. HyperMap

HyperMap [7] is a method to map a network into its hyperbolic space based on Maximum Likelihood Estimation. Unlike the coalescent embedding techniques, it can only perform the embedding in two dimensions and cannot exploit the information of the weights. It replays the hyperbolic growth of the network and at each step it finds the polar coordinates of the added node by maximizing the likelihood that the network was produced by the E-PSO model [7].

For curvature $K = -1$ the procedure is as follows: (1) Nodes are sorted decreasingly by degree and then labeled $i = 1, 2, \ldots, N$ according to the order; (2) Node $i = 1$ is born and assigned radial coordinate $r_1 = 0$ and a random angular coordinate $\theta_1 \in [0, 2\pi]$; (3) For each node $i = 2, 3, \ldots, N$ do: (3.a) Node $i$ is added to the network and assigned a radial coordinate $r_i = 2 \ln i$; (3.b) The radial coordinate of every existing node $j < i$ is increased according to $r_j(i) = \beta r_j + (1 - \beta) r_i$; (3.c) The node $i$ is assigned an angular coordinate by maximizing the likelihood $L_i = \prod_{1 \leq j < i} p(h_{ij})^{x_{ij}} (1 - p(h_{ij}))^{1 - x_{ij}}$, where $\beta \in (0, 1]$ is obtained from the exponent $\gamma = 1 + 1/\beta$ of the power law degree distribution, $p(h_{ij})$ is the connection probability of nodes $i$ and $j$ at hyperbolic distance $h_{ij}$ [7] and $x_{ij}$ is the adjacency matrix. The maximization is done by numerically trying different angular coordinates in steps of $2\pi/N$ and choosing the one that leads to the biggest $L_i$. The method has been implemented in MATLAB.

Given the coordinates of the embedding, the link prediction using HD and HSP are performed as described in the coalescent embedding section 1.4.

### 1.6. HyperMap-CN

HyperMap-CN [9] is a further development of HyperMap, where the inference of the angular coordinates is not performed anymore maximizing the likelihood $L_{i,L}$, based on the connections and disconnections of the nodes, but using another local likelihood $L_{i,CN}$, based on the number of common neighbours between each node $i$ and the previous nodes $j < i$ at final time. Here the hybrid model has been used, a variant of the method in which the likelihood $L_{i,CN}$ is only adopted for the high degree nodes and $L_{i,L}$ for the others, yielding a shorter running time. Furthermore, a speed-up heuristic and corrections steps can be applied. The speed-up can be achieved by getting an initial estimate of the angular coordinate of a node $i$ only considering the previous nodes $j < i$ that are $i$'s neighbours. The maximum likelihood estimation is then performed only looking at an interval around this initial estimate. Correction steps can be used at predefined times $i$ after step 3.c (in the description of HyperMap). Each existing node $j < i$ is visited and with the knowledge of the rest of the coordinates the angle of $j$ is updated to the value that maximizes the likelihood $L_{j,L}$. The C++ implementation of the method has been released by the authors at the website https://bitbucket.org/dk-lab/2015_code_hypermap. The default settings have been used (correction steps but no speed-up heuristic).

Given the coordinates of the embedding, the link prediction using HD and HSP are performed as described in the coalescent embedding section 1.4.

### 1.7. LPCS

Link Prediction with Community Structure (LPCS) [10] is a hyperbolic embedding technique that consists of the following steps: (1) Detect the hierarchical organization of communities. (2) Order the top-level communities starting from the one that has the largest number of nodes and using the Community Intimacy index, which takes into account the proportion of edges within and between communities. (3) Recursively order the lower level communities based on the order of the higher-level communities, until reaching the bottom level in the hierarchy. (4) Assign to every bottom-level community an angular range of size proportional to the nodes in the community, in order to cover the complete circle with non-overlapping angular ranges. Sample the angular coordinates of the nodes uniformly at random within the angular range of the related bottom-level community. (5) Assign the radial coordinates as described in the next paragraph.

The LPCS code firstly takes advantage of the Louvain R function for detecting the hierarchy of communities (see Louvain method), then we implemented the embedding in MATLAB. Given the coordinates of the embedding, the link prediction using HD and HSP are performed as described in the coalescent embedding section 1.4.

## 2. Community detection methods

### 2.1. Louvain

The Louvain algorithm [11] is separated into two phases, which are repeated iteratively.

At first every node in the (weighted) network represents a community in itself. In the first phase, for each node $i$, it considers its neighbours $j$ and evaluates the gain in modularity that would take place by removing $i$ from its community and placing it in the community of $j$. The node $i$ is then placed in the community $j$ for which this gain is maximum, but only if the gain is positive. If no gain is possible node $i$ stays in its original community. This process is applied until no further improvement can be achieved.

In the second phase the algorithm builds a new network whose nodes are the communities found in the first phase, whereas the weights of the links between the new nodes are given by the sum of the weight of the links between nodes in the corresponding two communities. Links between nodes of the same community lead to self-loops for this community in the new network.

Once the new network has been built, the two phase process is iterated until there are no more changes and a maximum of modularity has been obtained. The number of iterations determines the height of the hierarchy of communities detected by the algorithm.

For each hierarchical level there is a possible partition to compare to the ground truth annotation. In this case, the hierarchical level considered is the one that guarantees the best match, therefore the detected partition that gives the highest NMI value.

We used the R function *multilevel.community*, an implementation of the method available in the *igraph* package[12].

### 2.2. Infomap

The Infomap algorithm [13] finds the community structure by minimizing the expected description length of a random walker trajectory using the Huffman coding process[14].

It uses the hierarchical map equation (a further development of the map equation, to detect community structures on more than one level) in the form $L(M) = q_\curvearrowright H(Q) + \sum_{i=1}^{m} L(M^i)$.

$L(M)$ is the lower bound of the code length to specify a network path of a partitioning M, $q_\curvearrowright H(Q)$ is the Shannon information at the coarsest level of the partitioning, $L(M^i) = q_\circlearrowleft^i H(Q^i) + \sum_{j=1}^{m^i} L(M^{ij})$ is the lower bound of the code length to specify a network path of a partitioning M at sublevel $i$ and $L(M^{ij...k}) = p_{\circlearrowright}^{ij...k} H(P^{ij...k})$ is the lower bound of the code length at the finest modular level with submap $M^{ij...k}$.

The hierarchical map equation indicates the theoretical limit of how concisely a network path can be specified using a given partition structure. In order to calculate the optimal partition (community) structure, this limit can be computed for different partitions and the community annotation that gives the shortest path length is chosen.

For each hierarchical level there is a possible partition to compare to the ground truth annotation. In this case, the hierarchical level considered is the one that guarantees the best match, therefore the detected partition that gives the highest NMI value.

We used the C implementation released by the authors at http://www.mapequation.org/code.html.

### 2.3. Walktrap

The Walktrap algorithm [15] is based on an agglomerative method for hierarchical clustering: the nodes are iteratively grouped into communities exploiting the similarities between them. The nodes similarities are obtained using random walks and are based on the idea that random walks tend to get trapped into densely connected subgraphs corresponding to communities.

The agglomerative method uses heuristics to choose which communities to merge and implements an efficient way to update the distances between communities. At the end of the procedure a hierarchy of communities is obtained and each level offers a possible partition. The algorithm chooses as final result the partition that maximizes the modularity.

We used the R function *walktrap.community*, an implementation of the method available in the *igraph* package[12].

### 2.4. Label propagation

The label propagation algorithm [16] initializes each node with a unique label and iteratively updates each node label with the one owned by the majority of the neighbours, with ties broken uniformly at random. The update is performed in an asynchronous way and the order of the nodes at each iteration is chosen randomly. As the labels propagate through the network, densely connected groups of nodes quickly reach a consensus on a unique label. The iterative process stops when every node has the same label as the majority its neighbours, ties included. At the end of the procedure the nodes having the same label are grouped together to form a community. Since the aim is not the optimization of an objective function and the propagation process contains randomness, there are more possible partitions that satisfy the stop criterion and therefore the solution is not unique.

We used the R function *label.propagation.community*, an implementation of the method available in the *igraph* package[12].

## 2.5. Normalized Mutual Information

The evaluation of the community detection has been performed using the Normalized Mutual Information (NMI) as in [17]. The entropy can be defined as the information contained in a distribution *p(x)* in the following way:

$$H(X) = \sum_{x \in X} p(x) \log p(x)$$

The mutual information is the shared information between two distributions:

$$I(X,Y) = \sum_{y \in Y} \sum_{x \in X} p(x,y) \log \left( \frac{p(x,y)}{p_1(x) p_2(y)} \right)$$

To normalize the value between 0 and 1 the following formula can be applied:

$$NMI = \frac{I(X,Y)}{\sqrt{H(X)H(Y)}}$$

If we consider a partition of the nodes in communities as a distribution (probability of one node falling into one community), we can compute the matching between the annotation obtained by the community detection algorithm and the ground truth communities of a network as follows:

$$H(C_D) = \sum_{h=1}^{n_D} \frac{n_h^D}{N} \log(\frac{n_h^D}{N})$$

$$H(C_T) = \sum_{l=1}^{n_T} \frac{n_l^T}{N} \log(\frac{n_l^T}{N})$$

$$I(C_D, C_T) = \sum_h \sum_l \frac{n_{h,l}}{N} \log \left( \frac{n_{h,l}}{n_h^D n_l^T} \right).$$

$$NMI(C_D, C_T) = \frac{I(C_D, C_T)}{\sqrt{H(C_D)H(C_T)}}$$

Where:

N – number of nodes;

$n^D, n^T$ – number of communities detected by the algorithm (D) or ground truth (T);

$n_{h,l}$ – number of nodes assigned to the *h*-th community by the algorithm and to the *l*-th

community according to the ground truth annotation.

We used the MATLAB implementation available at http://commdetect.weebly.com/. As suggested in the code, when $\frac{N}{n^T} \leq 100$, the NMI should be adjusted in order to correct for chance [18].

## 3. Real networks datasets

The real networks have been transformed into undirected and unweighted, self-loops have been removed and the largest connected component has been considered.

*Mouse neural*: in-vivo single neuron connectome that reports mouse primary visual cortex (layers 1, 2/3 and upper 4) synaptic connections between neurons [19].

*Karate*: social network of a university karate club collected by Wayne Zachary in 1977. Each node represents a member of the club and each edge represents a tie between two members of the club [20].

*Dolphins*: a social network of bottlenose dolphins. The nodes are the bottlenose dolphins (genus Tursiops) of a bottlenose dolphin community living off Doubtful Sound, a fjord in New Zealand. An edge indicates a frequent association. The dolphins were observed between 1994 and 2001 [21].

*Macaque neural*: a macaque cortical connectome, assembled in previous studies in order to merge partial information obtained from disparate literature and database sources [22].

*Polbooks*: nodes represent books about US politics sold by the online bookseller Amazon.com. Edges represent frequent co-purchasing of books by the same buyers, as indicated by the "customers who bought this book also bought these other books" feature on Amazon. The network was compiled by V. Krebs and is unpublished, but can found at http://www-personal.umich.edu/~mejn/netdata/.

*ACM2009_contacts*: network of face-to-face contacts (active for at least 20 seconds) of the attendees of the ACM Conference on Hypertext and Hypermedia 2009 [23].

*Football*: network of American football games between Division IA colleges during regular season Fall 2000 [24].

*Physicians innovation*: the network captures innovation spread among physicians in the towns in Illinois, Peoria, Bloomington, Quincy and Galesburg. The data was collected in 1966. A node represents a physician and an edge between two physicians shows that the left physician told that the right physician is his friend or that he turns to the right physician if he needs advice or is interested in a discussion [25].

*Manufacturing email*: email communication network between employees of a mid-sized manufacturing company [26].

*Littlerock foodweb*: food web of Little Rock Lake, Wisconsin in the United States of America. Nodes are autotrophs, herbivores, carnivores and decomposers; links represent food sources [27].

*Jazz*: collaboration network between Jazz musicians. Each node is a Jazz musician and an edge denotes that two musicians have played together in a band. The data was collected in 2003 [28].

*Residence hall friends*: friendship network between residents living at a residence hall located on the Australian National University campus [29].

*Haggle contacts*: contacts between people measured by carried wireless devices. A node represents a person and an edge between two persons shows that there was a contact between them [30].

*Worm nervous*: a C. *Elegans* connectome representing synaptic interactions between neurons [31].

*Netsci*: a co-authorship network of scientists working on networks science [32].

*Infectious contacts*: network of face-to-face contacts (active for at least 20 seconds) of people during the exhibition INFECTIOUS: STAY AWAY in 2009 at the Science Gallery in Dublin [23].

*Flightmap*: a network of flights between American and Canadian cities [33].

*Email*: email communication network at the University Rovira i Virgili in Tarragona in the south of Catalonia in Spain. Nodes are users and each edge represents that at least one email was sent [34].

*Polblog*: a network of front-page hyperlinks between blogs in the context of the 2004 US election. A node represents a blog and an edge represents a hyperlink between two blogs [35].

*Odlis*: Online Dictionary of Library and Information Science (ODLIS): ODLIS is designed to be a hypertext reference resource for library and information science professionals, university students and faculty, and users of all types of libraries. Version December 2000 [36].

*Advogato*: a trust network of the online community platform Advogato for developers of free software launched in 1999. Nodes are users of Advogato and the edges represent trust relationships [37].

*Arxiv astroph*: collaboration graph of authors of scientific papers from the arXiv's Astrophysics (astro-ph) section. An edge between two authors represents a common publication [38].

*Thesaurus*: this is the Edinburgh Associative Thesaurus. Nodes are English words, and a directed link from A to B denotes that the word B was given as a response to the stimulus word

A in user experiments [39].

*Arxiv hepth*: this is the network of publications in the arXiv's High Energy Physics – Theory (hep-th) section. The links that connect the publications are citations [38].

*Facebook*: a network of a small subset of posts to user's walls on Facebook. The nodes of the network are Facebook users, and each edge represents one post, linking the users writing a post to the users whose wall the post is written on [40].

*ARK200909-ARK201012*: Autonomous systems (AS) Internet topologies extracted from the data collected by the Archipelago active measurement infrastructure (ARK) developed by CAIDA, from September 2009 up to December 2010. The connections in the topology are not physical but logical, representing AS relationships [41].

Most of the networks in the dataset can be downloaded from the Koblenz Network Collection at http://konect.uni-koblenz.de.

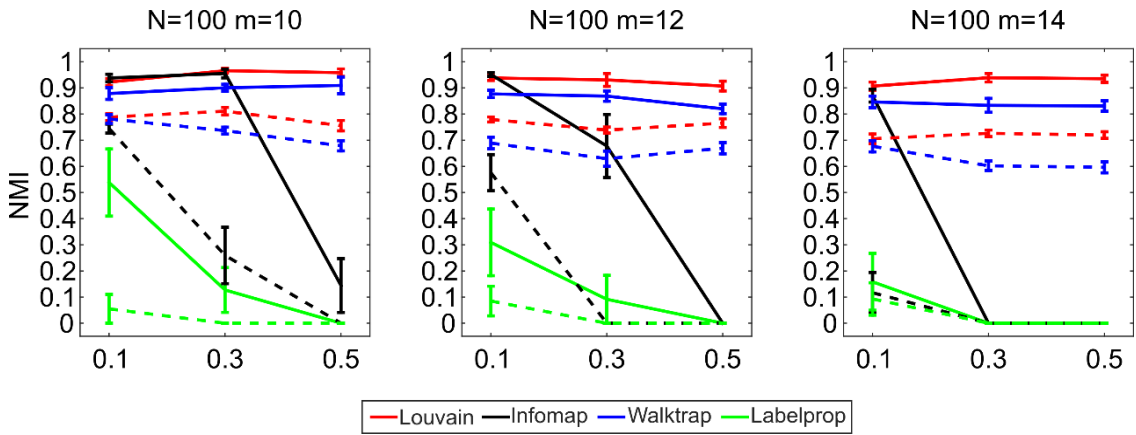

**Figure 1. Comparison of performance on small nPSO networks for 4 and 8 communities.**
The same results reported in Figure 3-4 in the main article for networks with $N$=100 are here shown in a unique plot, in order to highlight the decrease of performance of the community detection methods when the number of communities is increased from 4 (full lines) to 8 (dashed lines).

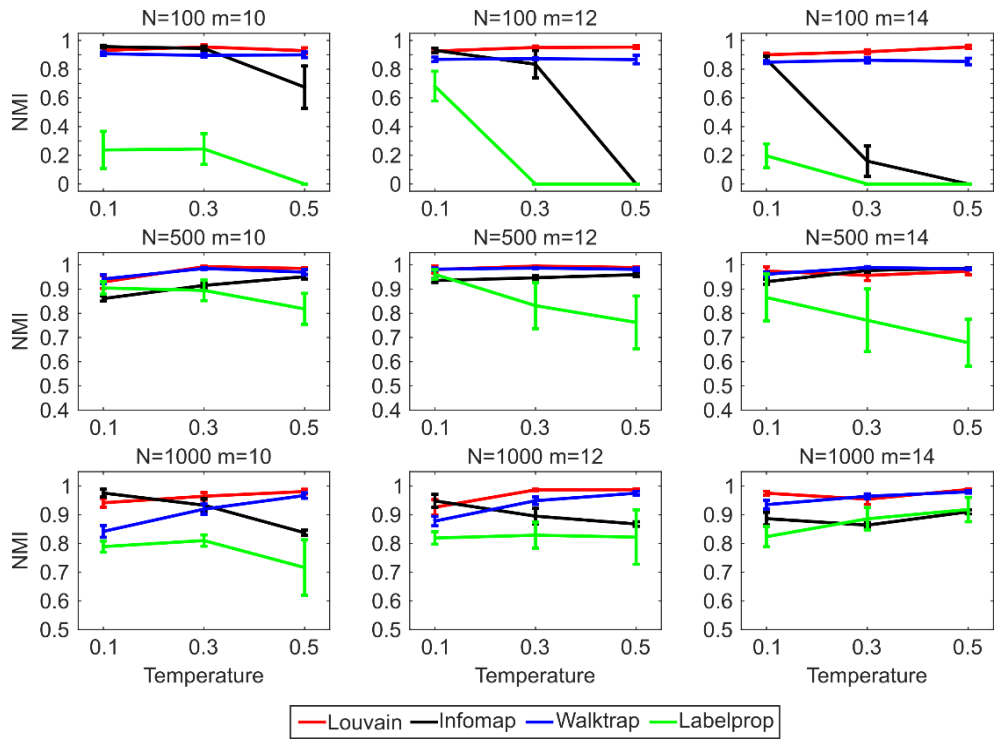

**Figure 2. Community detection on nPSO networks with 4 communities: implementation 2.**
The figure is equivalent to Figure 3 in the main article, but the networks are generated using the implementation 2. The figure is provided in order to support the equivalence of the community detection results using different implementations of the model.

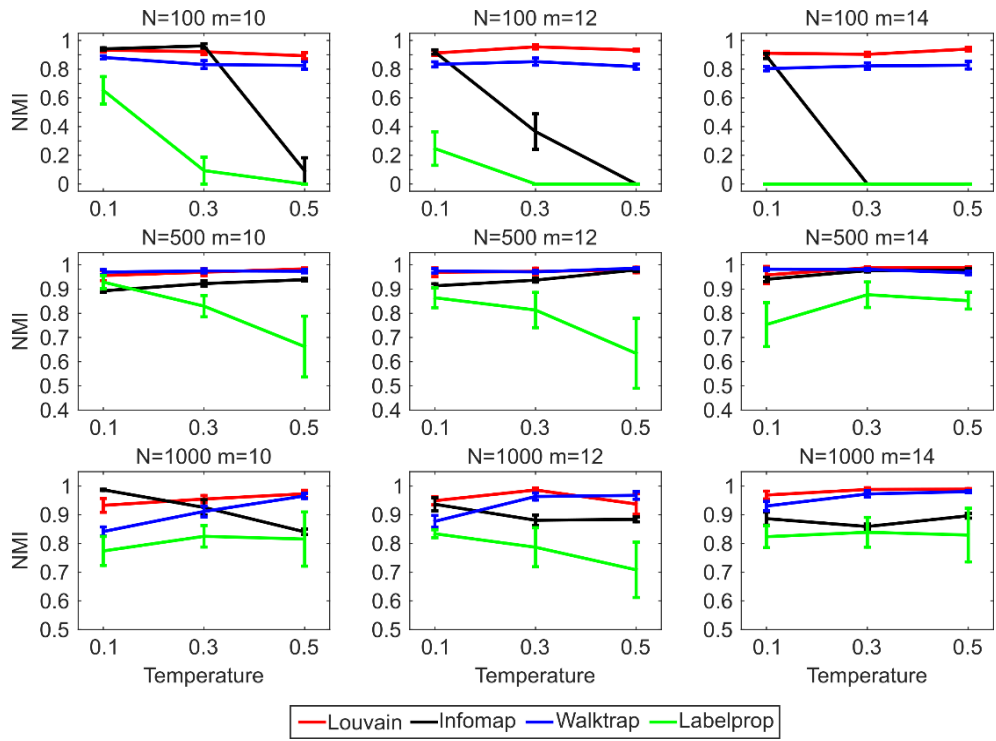

**Figure 3. Community detection on nPSO networks with 4 communities: implementation 3.**
The figure is equivalent to Figure 3 in the main article, but the networks are generated using the implementation 3. The figure is provided in order to support the equivalence of the community detection results using different implementations of the model.

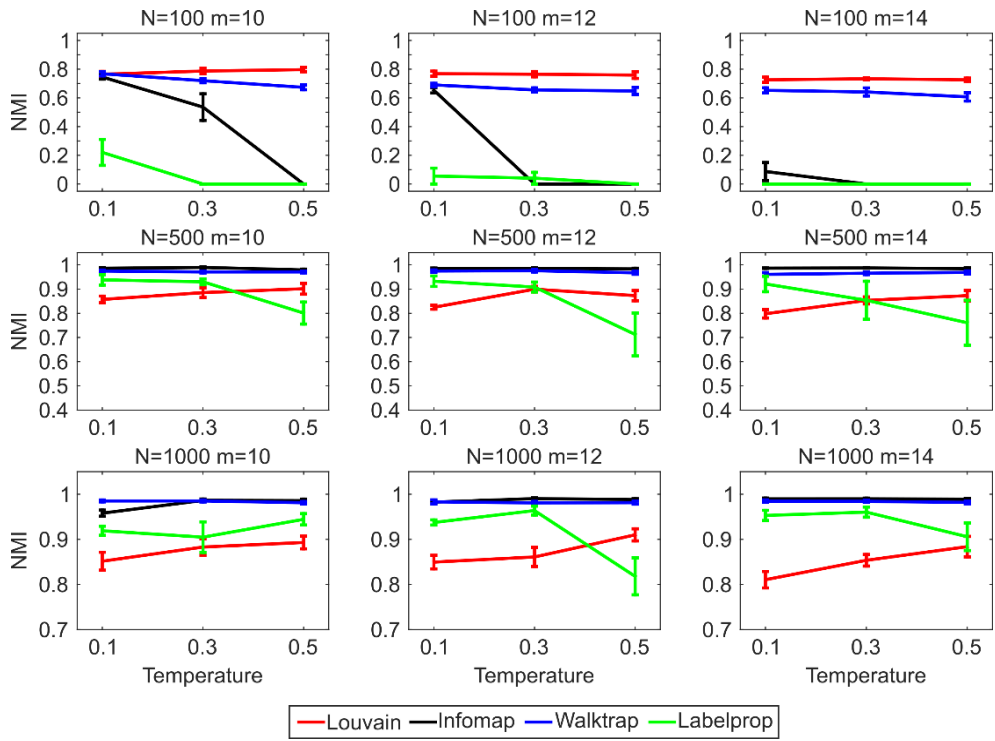

**Figure 4. Community detection on nPSO networks with 8 communities: implementation 2.**
The figure is equivalent to Figure 4 in the main article, but the networks are generated using the implementation 2. The figure is provided in order to support the equivalence of the community detection results using different implementations of the model.

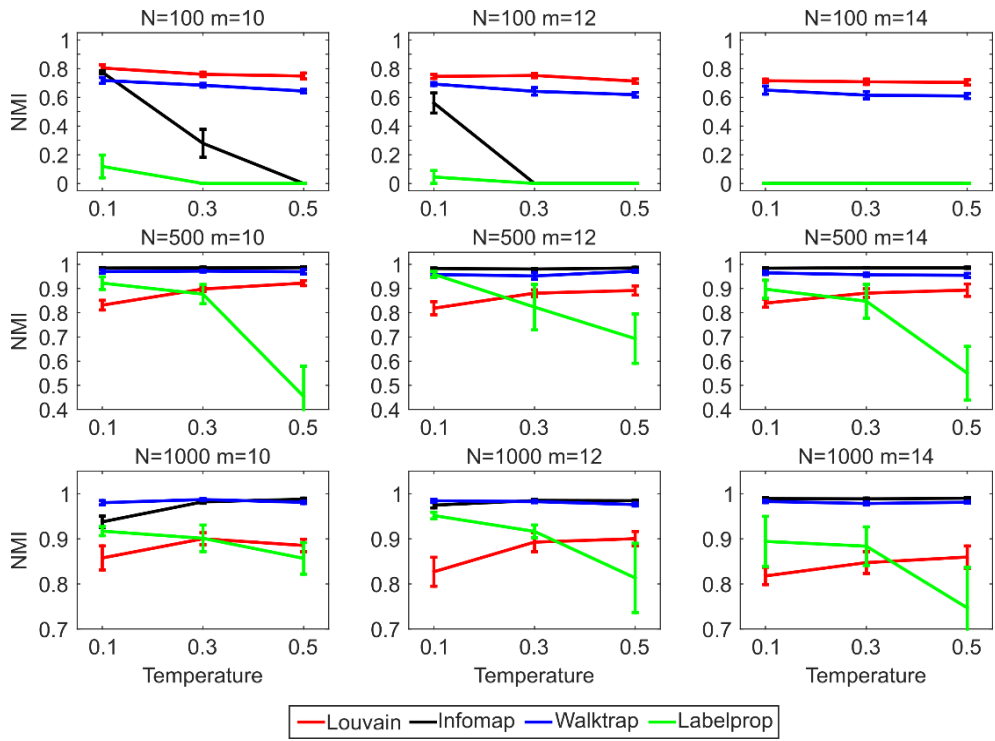

**Figure 5. Community detection on nPSO networks with 8 communities: implementation 3.**
The figure is equivalent to Figure 4 in the main article, but the networks are generated using the implementation 3. The figure is provided in order to support the equivalence of the community detection results using different implementations of the model.

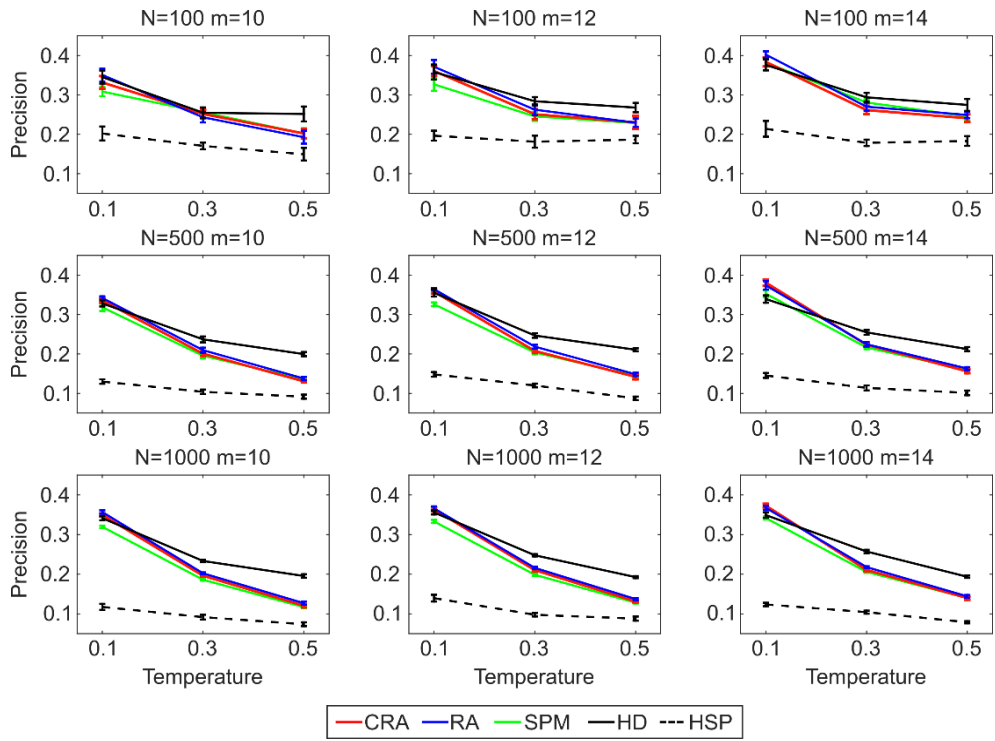

**Figure 6. Link prediction on PSO networks: implementation 2.**
The figure is equivalent to Figure 5 in the main article, but the networks are generated using the implementation 2. The figure is provided in order to support the equivalence of the link prediction results using different implementations of the model.

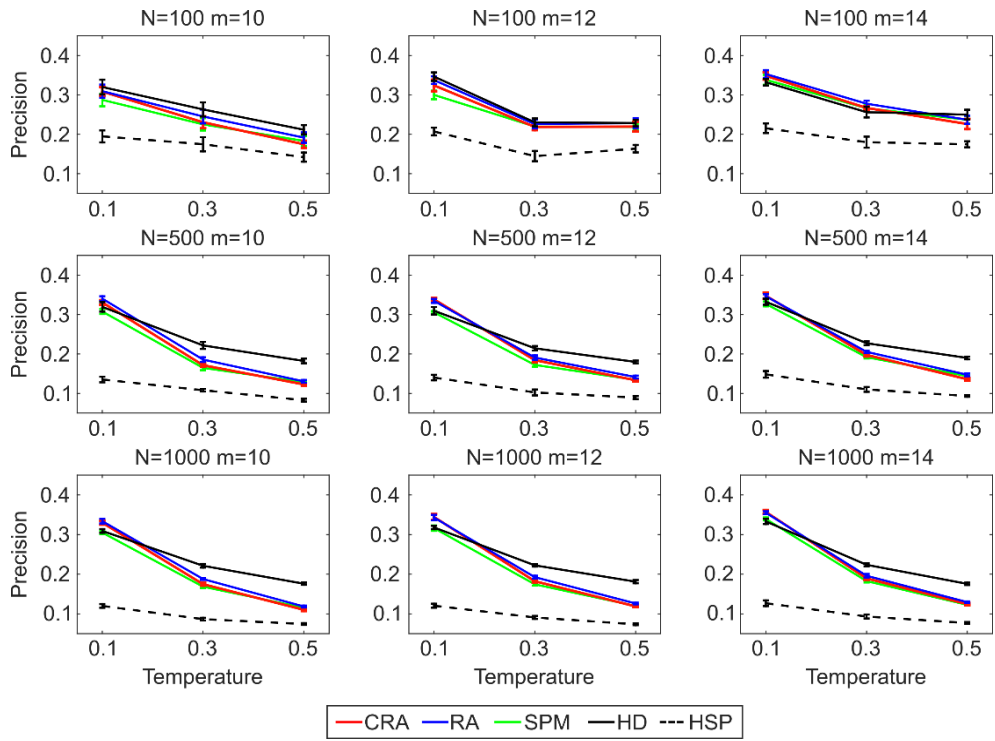

**Figure 7. Link prediction on PSO networks: implementation 3.**
The figure is equivalent to Figure 5 in the main article, but the networks are generated using the implementation 3. The figure is provided in order to support the equivalence of the link prediction results using different implementations of the model.

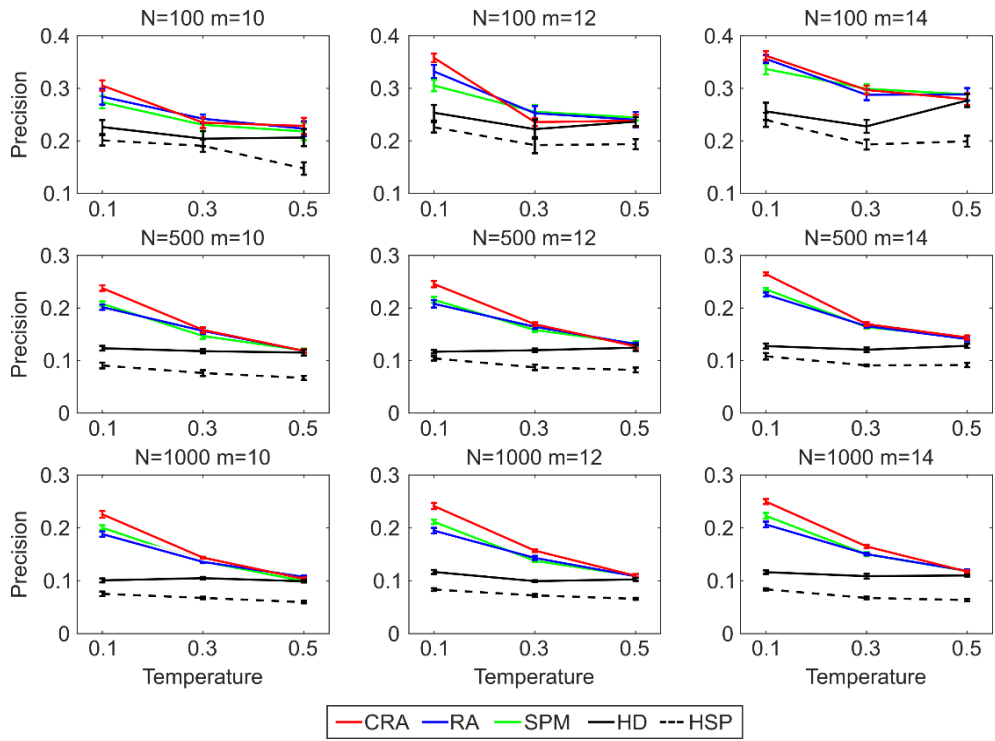

**Figure 8. Link prediction on nPSO networks with 4 communities: implementation 2.**
The figure is equivalent to Figure 6 in the main article, but the networks are generated using the implementation 2. The figure is provided in order to support the equivalence of the link prediction results using different implementations of the model.

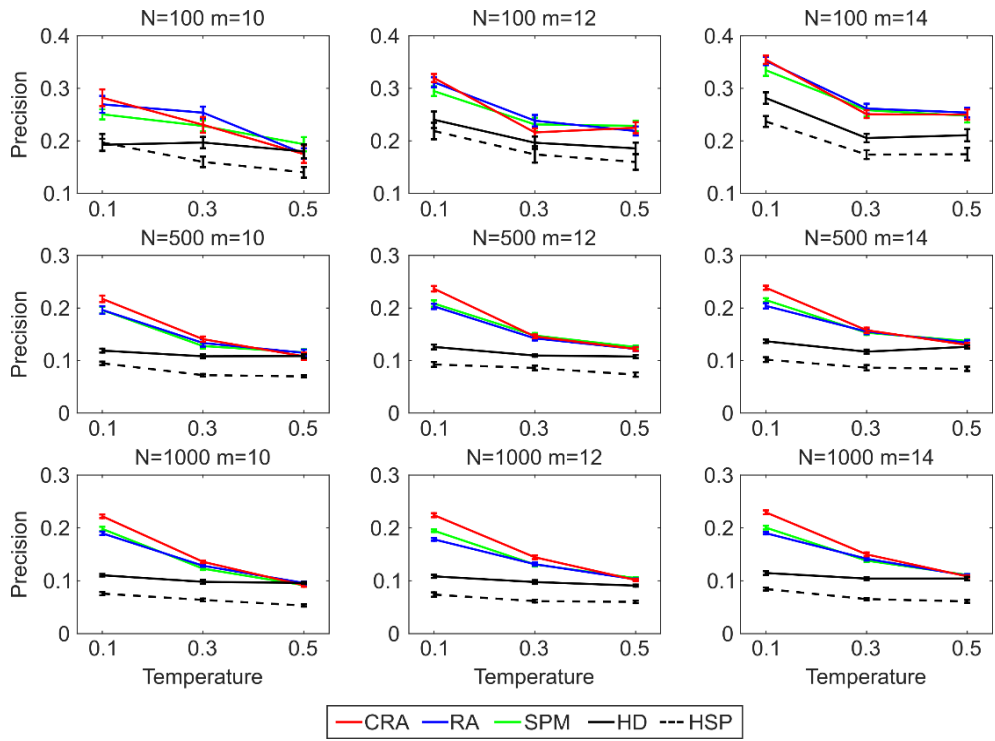

**Figure 9. Link prediction on nPSO networks with 4 communities: implementation 3.**
The figure is equivalent to Figure 6 in the main article, but the networks are generated using the implementation 3. The figure is provided in order to support the equivalence of the link prediction results using different implementations of the model.

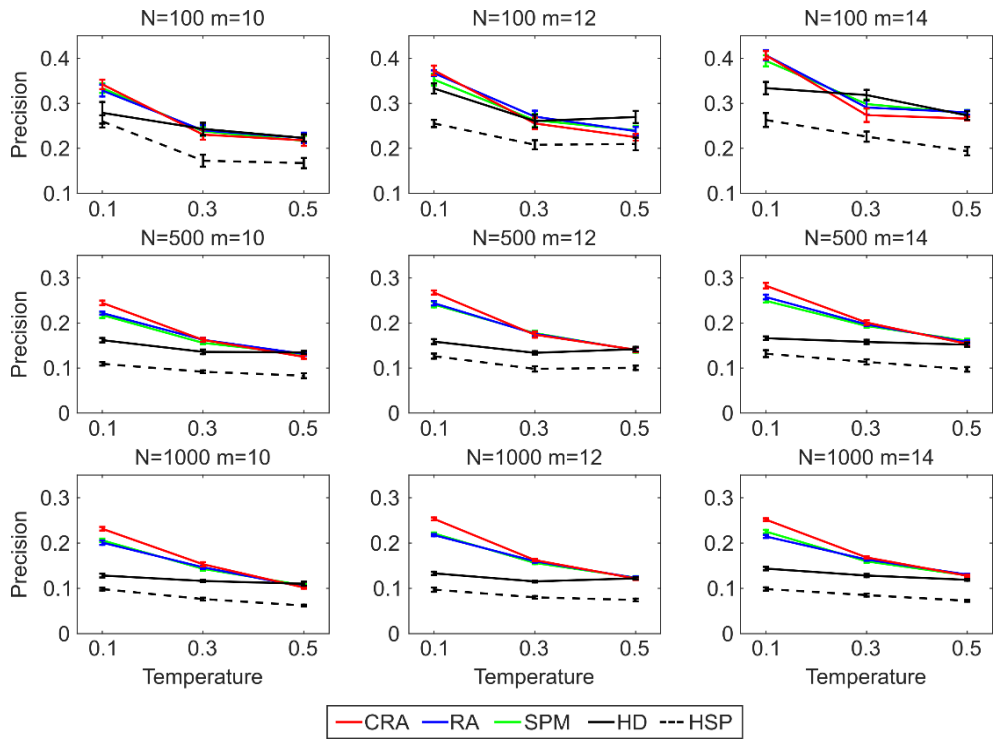

**Figure 10. Link prediction on nPSO networks with 8 communities: implementation 2.**
The figure is equivalent to Figure 7 in the main article, but the networks are generated using the implementation 2. The figure is provided in order to support the equivalence of the link prediction results using different implementations of the model.

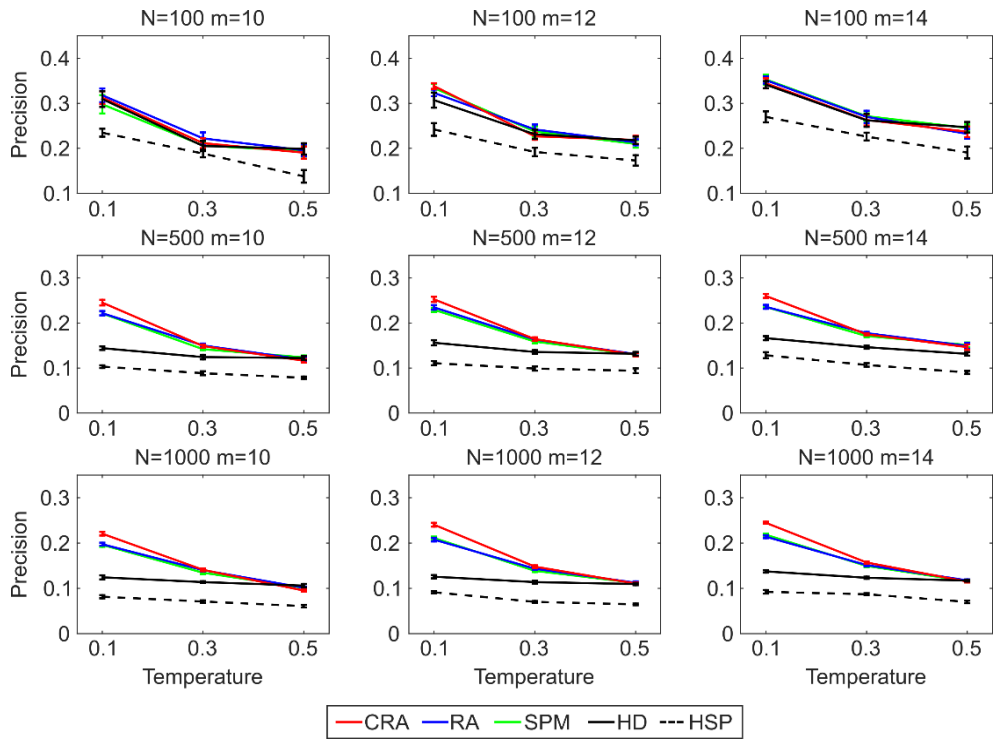

**Figure 11. Link prediction on nPSO networks with 8 communities: implementation 3.**
The figure is equivalent to Figure 7 in the main article, but the networks are generated using the implementation 3. The figure is provided in order to support the equivalence of the link prediction results using different implementations of the model.

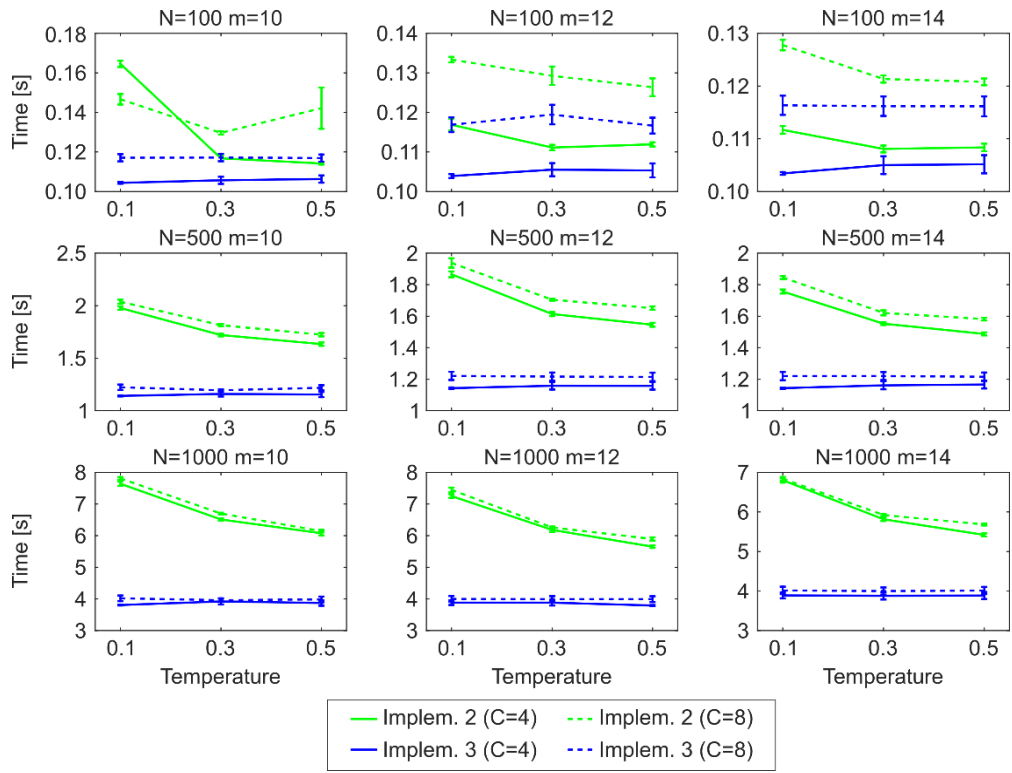

**Figure 12. Time performance for generating nPSO networks: implementations 2-3.**
Due to the different scale of the computational time of implementation 1 reported in Figure 9 and 10 in the main article, the figure reports the time performance for generating nPSO networks with 4 and 8 communities only for implementations 2 and 3, allowing a more detailed comparison.

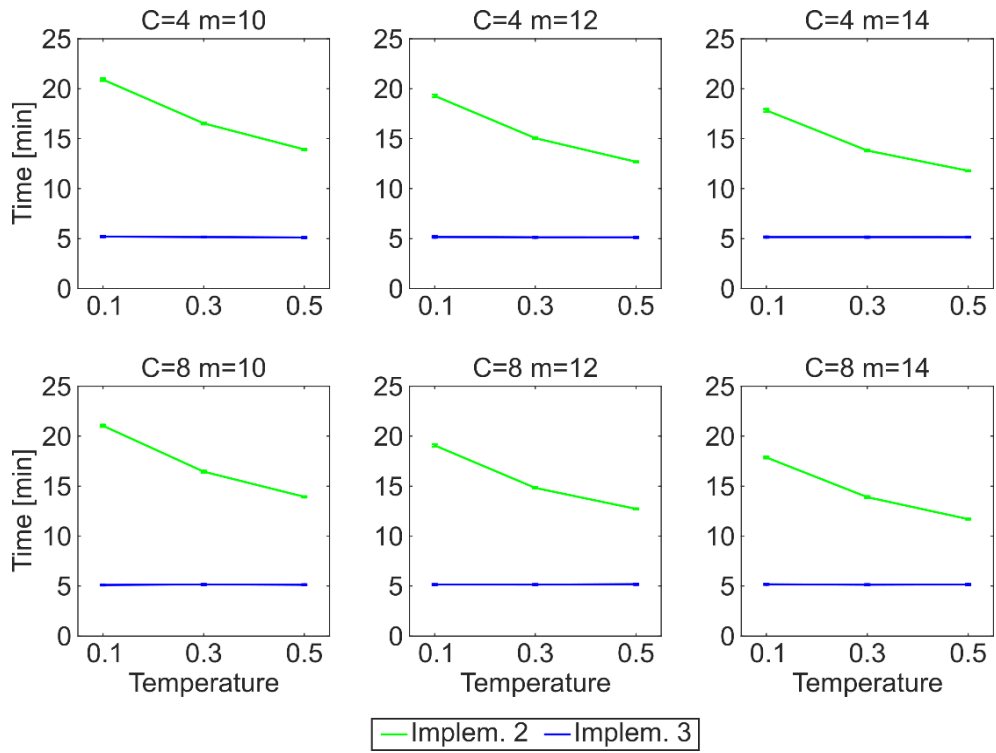

**Figure 13. Time performance for generating large-size nPSO networks.**
Synthetic networks have been generated using the nPSO model with parameters $\gamma = 3$ (power-law degree distribution exponent), $m = [10, 12, 14]$ (half of average degree), $T = [0.1, 0.3, 0.5]$ (temperature, inversely related to the clustering coefficient), $N = 10000$ (network size) and $C = [4, 8]$ (communities). For each combination of parameters, 10 networks have been generated using the implementations 2 and 3. The plots report for each parameter combination the mean computational time and standard error over the random iterations.

**Table 1. Statistics of small-size real networks.**
For each network several statistics have been computed. *N* is the number of nodes. *E* is the number of edges. The parameter *m*, as in the PSO model, refers to half of the average node degree. *D* is the network density. *C* is the average clustering coefficient, computed for each node as the number of links between its neighbours over the number of possible links [31]. *L* is the characteristic path length of the network [31]. Power-law is the exponent $\gamma$ of the power-law distribution estimated from the observed degree distribution of the network using the maximum likelihood procedure described in [8].

|  | N | E | m | D | C | L | Power law |
|---|---|---|---|---|---|---|---|
| mouse neural | 18 | 37 | 2.06 | 0.24 | 0.22 | 1.97 | 4.01 |
| karate | 34 | 78 | 2.29 | 0.14 | 0.57 | 2.41 | 2.12 |
| dolphins | 62 | 159 | 2.56 | 0.08 | 0.26 | 3.36 | 6.96 |
| macaque neural | 94 | 1515 | 16.12 | 0.35 | 0.77 | 1.77 | 4.46 |
| polbooks | 105 | 441 | 4.20 | 0.08 | 0.49 | 3.08 | 2.62 |
| ACM2009 contacts | 113 | 2196 | 19.43 | 0.35 | 0.53 | 1.66 | 3.74 |
| football | 115 | 613 | 5.33 | 0.09 | 0.40 | 2.51 | 9.09 |
| physicians innovation | 117 | 465 | 3.97 | 0.07 | 0.22 | 2.59 | 4.51 |
| manufacturing email | 167 | 3250 | 19.46 | 0.23 | 0.59 | 1.97 | 3.13 |
| littlerock foodweb | 183 | 2434 | 13.30 | 0.15 | 0.32 | 2.15 | 3.00 |
| jazz | 198 | 2742 | 13.85 | 0.14 | 0.62 | 2.24 | 4.48 |
| residence hall friends | 217 | 1839 | 8.47 | 0.08 | 0.36 | 2.39 | 6.32 |
| haggle contacts | 274 | 2124 | 7.75 | 0.06 | 0.63 | 2.42 | 1.51 |
| worm nervoussys | 297 | 2148 | 7.23 | 0.05 | 0.29 | 2.46 | 3.34 |
| netsci | 379 | 914 | 2.41 | 0.01 | 0.74 | 6.04 | 3.36 |
| infectious contacts | 410 | 2765 | 6.74 | 0.03 | 0.46 | 3.63 | 6.42 |
| flightmap | 456 | 37947 | 83.22 | 0.37 | 0.81 | 1.64 | 1.71 |
| email | 1133 | 5451 | 4.81 | 0.01 | 0.22 | 3.61 | 4.89 |
| polblog | 1222 | 16714 | 13.68 | 0.02 | 0.32 | 2.74 | 2.38 |

**Table 2. Statistics of AS Internet snapshots and large-size real networks.**
The first half of the table reports the statistics for the AS Internet snapshots, whereas the second half the large-size real networks. Note that also the last AS Internet snapshot has been considered in the simulations on the large-size real networks. For each network several statistics have been computed. $N$ is the number of nodes. $E$ is the number of edges. The parameter $m$, as in the PSO model, refers to half of the average node degree. $D$ is the network density. $C$ is the average clustering coefficient, computed for each node as the number of links between its neighbours over the number of possible links [31]. $L$ is the characteristic path length of the network [31]. Power-law is the exponent $\gamma$ of the power-law distribution estimated from the observed degree distribution of the network using the maximum likelihood procedure described in [8].

|  | N | E | m | D | C | L | Power law |
|---|---|---|---|---|---|---|---|
| ARK200909 | 24091 | 59531 | 2.47 | 0.0039 | 0.36 | 3.53 | 2.12 |
| ARK200912 | 25910 | 63435 | 2.45 | 0.0031 | 0.36 | 3.54 | 2.11 |
| ARK201003 | 26307 | 66089 | 2.51 | 0.0012 | 0.37 | 3.53 | 2.26 |
| ARK201006 | 26756 | 68150 | 2.55 | 0.0011 | 0.37 | 3.51 | 2.08 |
| ARK201009 | 28353 | 73722 | 2.60 | 0.0009 | 0.37 | 3.52 | 2.23 |
| ARK201012 | 29333 | 78054 | 2.66 | 0.0002 | 0.38 | 3.50 | 2.22 |
| odlis | 2898 | 16376 | 5.65 | 0.0002 | 0.30 | 3.17 | 2.63 |
| advogato | 5042 | 39227 | 7.78 | 0.0002 | 0.25 | 3.27 | 2.73 |
| arxiv astroph | 17903 | 196972 | 11.00 | 0.0002 | 0.63 | 4.19 | 2.83 |
| thesaurus | 23132 | 297094 | 12.84 | 0.0002 | 0.09 | 3.49 | 2.84 |
| arxiv hepth | 27400 | 352021 | 12.85 | 0.0002 | 0.31 | 4.28 | 2.86 |
| facebook | 43953 | 182384 | 4.15 | 0.0002 | 0.11 | 5.60 | 3.66 |

**Table 3. Clustering coefficient comparison for different implementations.**
Synthetic networks have been generated using both the PSO and nPSO (4 and 8 communities) model, with parameters $\gamma = 3$ (power-law degree distribution exponent), $m = [10, 12, 14]$ (half of average degree), $T = [0.1, 0.3, 0.5]$ (temperature, inversely related to the clustering coefficient) and $N = [100, 500, 1000]$ (network size). For each combination of parameters, 10 networks have been generated using the 3 different implementations and their average clustering coefficient has been computed [31]. The table reports for each combination of parameters the mean clustering coefficient over the 10 networks generated using the implementation 1. Instead, for the implementations 2 and 3, the difference in the mean clustering coefficient with respect to the implementation 1 is reported.

|   |   |   | Clustering Implementation 1 | | | Clustering Difference Implementations 1-2 | | | Clustering Difference Implementations 1-3 | | |
|---|---|---|---|---|---|---|---|---|---|---|---|
|   |   |   | T=0.1 | T=0.3 | T=0.5 | T=0.1 | T=0.3 | T=0.5 | T=0.1 | T=0.3 | T=0.5 |
| PSO | N=100 | m=10 | 0.64 | 0.51 | 0.43 | -0.01 | -0.01 | 0.00 | 0.01 | 0.01 | 0.02 |
|  |  | m=12 | 0.66 | 0.53 | 0.45 | -0.01 | -0.01 | -0.01 | 0.00 | 0.01 | 0.01 |
|  |  | m=14 | 0.66 | 0.55 | 0.48 | -0.01 | -0.01 | -0.01 | 0.01 | 0.01 | 0.01 |
|  | N=500 | m=10 | 0.65 | 0.48 | 0.32 | 0.00 | -0.01 | 0.00 | 0.01 | 0.01 | 0.01 |
|  |  | m=12 | 0.65 | 0.48 | 0.33 | -0.01 | -0.01 | 0.00 | 0.01 | 0.01 | 0.01 |
|  |  | m=14 | 0.65 | 0.48 | 0.34 | 0.00 | -0.01 | 0.00 | 0.01 | 0.01 | 0.01 |
|  | N=1000 | m=10 | 0.65 | 0.48 | 0.31 | 0.00 | -0.01 | 0.00 | 0.01 | 0.01 | 0.01 |
|  |  | m=12 | 0.65 | 0.48 | 0.31 | 0.00 | 0.00 | 0.00 | 0.01 | 0.02 | 0.01 |
|  |  | m=14 | 0.66 | 0.49 | 0.32 | 0.00 | 0.00 | 0.00 | 0.01 | 0.02 | 0.01 |
| nPSO C=4 | N=100 | m=10 | 0.66 | 0.53 | 0.43 | -0.01 | -0.01 | -0.02 | 0.01 | 0.01 | 0.00 |
|  |  | m=12 | 0.67 | 0.55 | 0.46 | -0.01 | -0.01 | -0.01 | 0.01 | 0.01 | 0.01 |
|  |  | m=14 | 0.68 | 0.56 | 0.49 | -0.01 | -0.01 | -0.02 | 0.01 | 0.01 | 0.01 |
|  | N=500 | m=10 | 0.65 | 0.48 | 0.32 | -0.01 | -0.01 | -0.01 | 0.01 | 0.00 | 0.00 |
|  |  | m=12 | 0.66 | 0.49 | 0.33 | 0.00 | 0.00 | 0.00 | 0.01 | 0.01 | 0.01 |
|  |  | m=14 | 0.66 | 0.49 | 0.34 | 0.00 | 0.00 | 0.00 | 0.01 | 0.01 | 0.00 |
|  | N=1000 | m=10 | 0.65 | 0.48 | 0.31 | 0.00 | 0.00 | -0.01 | 0.00 | 0.01 | 0.00 |
|  |  | m=12 | 0.66 | 0.49 | 0.31 | -0.01 | -0.01 | 0.00 | 0.01 | 0.01 | 0.01 |
|  |  | m=14 | 0.66 | 0.49 | 0.32 | 0.00 | 0.00 | 0.00 | 0.01 | 0.01 | 0.00 |
| nPSO C=8 | N=100 | m=10 | 0.64 | 0.50 | 0.41 | -0.01 | -0.02 | 0.00 | -0.01 | 0.01 | 0.01 |
|  |  | m=12 | 0.65 | 0.52 | 0.44 | -0.01 | 0.00 | -0.01 | 0.00 | 0.02 | 0.01 |
|  |  | m=14 | 0.66 | 0.54 | 0.47 | 0.00 | -0.02 | -0.02 | 0.00 | 0.01 | 0.00 |
|  | N=500 | m=10 | 0.63 | 0.46 | 0.31 | 0.00 | 0.00 | 0.00 | 0.01 | 0.01 | 0.00 |
|  |  | m=12 | 0.64 | 0.47 | 0.32 | 0.00 | 0.00 | -0.01 | 0.01 | 0.01 | 0.01 |
|  |  | m=14 | 0.64 | 0.47 | 0.32 | 0.00 | -0.01 | -0.01 | 0.01 | 0.01 | 0.01 |
|  | N=1000 | m=10 | 0.64 | 0.47 | 0.30 | -0.01 | -0.01 | 0.00 | 0.00 | 0.01 | 0.01 |
|  |  | m=12 | 0.64 | 0.47 | 0.30 | 0.00 | 0.00 | 0.00 | 0.01 | 0.01 | 0.01 |
|  |  | m=14 | 0.65 | 0.47 | 0.30 | 0.00 | -0.01 | 0.00 | 0.00 | 0.01 | 0.01 |

|  | mean precision | mean ranking |
|---|---|---|
| EBC-ncISO-HSP | 0.17 | 9.9 |
| RA-LE-HSP | 0.17 | 11.6 |
| EBC-ncISO-EA-HSP | 0.16 | 12.1 |
| EBC-ISO-HSP | 0.15 | 13.4 |
| RA-ncISO-HSP | 0.16 | 13.4 |
| EBC-ISO-EA-HSP | 0.16 | 13.6 |
| EBC-LE-EA-HSP | 0.16 | 13.8 |
| EBC-MCE-HSP | 0.16 | 13.9 |
| EBC-ncMCE-HSP | 0.16 | 14.0 |
| HyperMapCN-HSP | 0.14 | 14.4 |
| EBC-MCE-EA-HSP | 0.16 | 14.6 |
| EBC-ncMCE-EA-HSP | 0.16 | 14.7 |
| RA-LE-EA-HSP | 0.15 | 14.8 |
| RA-MCE-HSP | 0.15 | 15.9 |
| RA-ncISO-EA-HSP | 0.15 | 16.0 |
| RA-ISO-HSP | 0.16 | 16.4 |
| RA-MCE-EA-HSP | 0.15 | 16.7 |
| RA-ncMCE-EA-HSP | 0.15 | 16.9 |
| RA-ncMCE-HSP | 0.14 | 17.0 |
| RA-ISO-EA-HSP | 0.14 | 18.8 |
| EBC-LE-HSP | 0.14 | 20.3 |
| HyperMapCN-HD | 0.11 | 20.3 |
| LPCS-HSP | 0.14 | 20.8 |
| HyperMap-HSP | 0.12 | 23.8 |
| EBC-ncISO-EA-HD | 0.12 | 25.8 |
| RA-LE-EA-HD | 0.12 | 27.3 |
| EBC-MCE-EA-HD | 0.12 | 27.6 |
| EBC-ncMCE-EA-HD | 0.12 | 28.1 |
| RA-ncMCE-EA-HD | 0.12 | 28.1 |
| RA-ncISO-EA-HD | 0.12 | 29.0 |
| LPCS-HD | 0.12 | 29.1 |
| HyperMap-HD | 0.10 | 29.7 |
| EBC-LE-EA-HD | 0.11 | 30.2 |
| EBC-ISO-EA-HD | 0.11 | 30.4 |
| EBC-ncISO-HD | 0.10 | 31.0 |
| RA-MCE-EA-HD | 0.11 | 31.5 |
| RA-ISO-EA-HD | 0.11 | 31.6 |
| RA-LE-HD | 0.09 | 33.3 |
| EBC-ncMCE-HD | 0.09 | 34.9 |
| EBC-MCE-HD | 0.08 | 35.3 |
| RA-ncISO-HD | 0.09 | 35.4 |
| RA-ISO-HD | 0.08 | 35.7 |
| EBC-ISO-HD | 0.07 | 36.6 |
| RA-ncMCE-HD | 0.08 | 37.2 |
| RA-MCE-HD | 0.07 | 37.8 |
| EBC-LE-HD | 0.07 | 38.2 |

**Table 4. Link prediction on small-size real networks using hyperbolic embedding methods.**
For each of the small-size real networks shown in Table 1, 10% of links have been randomly removed (10 iterations for HyperMap due to the high computational time, 100 iterations for the other methods) and the algorithms have been executed in order to assign likelihood scores to the non-observed links in these reduced networks. In order to evaluate the performance, the links are ranked by likelihood scores and the precision is computed as the percentage of removed links among the top-$r$ in the ranking, where $r$ is the total number of links removed. The table reports the mean precision and the mean ranking over the entire dataset. The methods are sorted by mean ranking.